\documentclass[twocolumn,pra,aps,showpacs]{revtex4}
\usepackage{amsmath}
\usepackage{graphicx}
\usepackage{amssymb}
\usepackage{amsthm}
\usepackage{txfonts}
\usepackage{color}
\usepackage[unicode=true,pdfusetitle,
 bookmarks=false,bookmarksnumbered=false,bookmarksopen=false,
 breaklinks=false,pdfborder={0 0 1},backref=false,colorlinks=false]
 {hyperref}

\newtheorem{thm}{Theorem}

\makeatother

\begin{document}

\title{No-go theorem and optimization of dynamical decoupling against noise
with soft cutoff}

\author{Zhen-Yu Wang}

\author{Ren-Bao Liu}

\email{rbliu@phy.cuhk.edu.hk}
\affiliation{Department of Physics and Center for Quantum Coherence, The Chinese
University of Hong Kong, Shatin, N. T., Hong Kong, China}

\begin{abstract}
We study the performance of dynamical decoupling in suppressing decoherence
caused by soft-cutoff Gaussian noise, using short-time expansion of
the noise correlations and numerical optimization.
For the noise with soft cutoff at high frequencies, there exists
no dynamical decoupling scheme to eliminate the decoherence to arbitrary
orders of the short time, regardless of the timing or pulse shaping
of the control under the population conserving condition. We formulate
the equations for optimizing pulse sequences that minimizes decoherence
up to the highest possible order of the short time
for the noise correlations with odd power terms in the short-time
expansion. In particular, we show that the Carr-Purcell-Meiboom-Gill
sequence is optimal in short-time limit for the noise correlations
with a linear order term in the time expansion.
\end{abstract}


\maketitle

\section{Introduction}

Quantum information processing~\cite{Nielsen:2000:Book} relies on
the coherence of quantum systems. Unavoidable interactions between
a quantum system and its environment (bath) introduce noise on the
system and lead to error evolutions (decoherence) of the quantum system.
Various methods have been proposed to combat the decoherence, including
decoherence-free subspaces~\cite{Duan:1997:DFS,Zanardi:1998:DFS,Lidar:1998:DFS},
error-correction codes~\cite{Shor:1995:QEC,Steane:1996:QEC}, and
dynamical decoupling (DD)~\cite{Viola:1998:PRA,Ban:1998:DD,Zanardi:1999:77,Viola:1999:2417,Yang:2010:2}.
In particular, the DD scheme uses rapid unitary control pulses acting
only on the systems to suppress the effects of the noise from the
environments. DD has the advantages of suppressing decoherence without
measurement, feedback, or redundant encoding~\cite{Viola:1999:2417}.
DD originated from the seminal spin echo experiment~\cite{Hahn:1950:Echo},
in which the effect of a static random magnetic field (inhomogeneous
broadening) is canceled. And more complex DD pulse sequences, such
as the Carr-Purcell-Meiboom-Gill (CPMG) sequence~\cite{Carr:1954:630,Meiboom:1958:688},
were designed to prolong the spin coherence time~\cite{Mehring:1983:BookNMR}.

The early DD schemes only eliminate low-order errors, i.e., the errors
of quantum evolutions up to some low order in the Magnus expansion.
By unitary symmetrization procedure~\cite{Zanardi:1999:77,Viola:1999:2417},
DD cancels the first order (i.e., leading order) errors. To eliminate
errors to the second order in short time, mirror-symmetric arrangement
of two DD sequences can be used~\cite{Viola:1999:2417}. The first
explicit arbitrary $M$th order DD scheme, which suppresses errors
to $\mathcal{O}(T^{M+1})$ for short evolution time $T$, is the concatenated
DD (CDD)~\cite{Khodjasteh:2005:180501,Khodjasteh:2007:062310} proposed
by Khodjasteh and Lidar. CDD sequences against pure dephasing were
investigated for electron spin qubits in realistic solid-state systems
with nuclear spins as baths~\cite{Yao:2007:077602,Witzel:2007:241303,Zhang:2007:201302}.
Experiments~\cite{Peng:2011:154003,Alvarez:2010:042306,Tyryshkin:1011.1903,Wang:1011.6417,Barthel:2010:266808}
have tested the performance of CDD. CDD works for generic quantum
systems coupled to a finite bath~\cite{Santos:2008:083009,Wang:2011:022306}.
However, since CDD uses recursively constructed pulse sequences to
suppress decoherence, the number of pulses increases exponentially
with the decoupling order. As pulse errors are inevitably introduced
in each control pulse in experiments, finding efficient DD schemes
with fewer control pulses is desirable. A remarkable advance is the
Uhrig DD (UDD)~\cite{Uhrig:2007:100504,Lee:2008:160505,Uhrig:2008:083024,Yang:2008:180403}.
UDD is optimal in the short-time limit in the sense that it suppresses
the pure dephasing of a qubit coupled to a finite bath to the $M$th
order using only $M$ qubit flips. The performance bounds for UDD
against pure dephasing were established~\cite{Uhrig:2010:012301}.
Shaped pulses~\cite{Pasini:2008:032315,Fauseweh:1112.0446} of finite
amplitude can be incorporated into UDD~\cite{Pasini:2008:032315}.
Many recent experimental studies~\cite{Biercuk:2009:996,Du:2009:1265,Michael:2009:062324,Uys:2009:040501,Jenista:2009:204510,Alvarez:2010:042306,Barthel:2010:266808}
demonstrated the performance of UDD.

It is important to find efficient schemes to suppress general decoherence
(including pure dephasing and population relaxation). Yang and Liu
extended UDD to the suppression of population relaxation~\cite{Yang:2008:180403}.
This inspired efficient ways to suppress the general decoherence of
single qubits, including concatenation of UDD sequences (CUDD)~\cite{Uhrig:2009:120502}
and a much more efficient one called quadratic DD (QDD)~\cite{West:2010:130501,Wang:2011:022306,Quiroz:1105.4303,Kuo:1106.2151,Jiang:2011:060302}
discovered by West \emph{et al}~\cite{West:2010:130501}. Based on
the proof in \cite{Yang:2008:180403}, Mukhtar \emph{et al} generalized
UDD to protect arbitrary multilevel systems with full prior knowledge
of the initial states~\cite{Mukhtar:2010:012331}. One can actually
preserve the coherence of arbitrary multi-qubit systems by protecting
a mutually orthogonal operation set (MOOS)~\cite{Wang:2011:022306}.
By nesting UDD sequences for protecting the elements in the MOOS,
the nested UDD (NUDD)~\cite{Wang:2011:022306,Mukhtar:2010:012331,Jiang:2011:060302}
requires only a polynomially increasing number of pulses in the decoupling
order. These universal DD schemes also work for analytically time-dependent
baths~\cite{Wang:2011:062313}.

The above-mentioned variations of UDD, however, rely on the finiteness
of the baths, i.e., the existence of hard high-frequency cutoff in
the noise spectra. Legitimate questions are: For quantum systems coupled
to an infinite quantum bath or affected by soft-cutoff noise, can
any DD be designed to eliminate the decoherence to arbitrary orders
of precision in the short-time limit? And if yes, how can such DD
be designed? Such questions have been previously addressed in some
specific noise models. Comparing the efficiency of various DD sequences
in suppressing pure dephasing of a qubit due to classical noise, Cywi\'{n}ski
\emph{et al} observed that if the noise spectrum cutoff is not reached,
CPMG sequences~\cite{Carr:1954:630,Meiboom:1958:688} actually performs
better than CDD and UDD sequences~\cite{Cywinski:2008:174509}.
With the consideration of minimum pulse separations in physical systems,
Viola \emph{ et al} observed that
low-order DD sequences provide better performance than high-order
DD when the rate of pulses is not faster than the correlation time
of the noise~\cite{Hodgson:2010:062321,Khodjasteh:2011:020305}.
It was confirmed by experiments that for $^{13}\text{C}$
spin qubits in a $^{1}\text{H}$ spin bath of which the high
frequency cutoff was not reached by the DD sequences, CPMG
outperforms UDD~\cite{Ajoy:2011:032303}.
Also, Pasini and Uhrig derived the equations for minimizing decoherence
for power-law spectra, and found that the numerically optimized sequences
resemble CPMG~\cite{Pasini:2010:012309}. Chen and Liu proved that
for telegraphlike noise the CPMG sequences are the most efficient
scheme in protecting the qubit coherence in the short-time limit and
the decoherence can be suppressed at most to the third order of short
evolution time by DD~\cite{Chen:2010:052324}. These results suggest
that for noise with soft cutoff in the spectrum, there are certain
constraints on the optimal order and decoupling scaling
of DD. Ref.~\cite{Gordon:2008:010403} presented numerical optimization of bounded-strength DD for specific noise spectra. However, no conclusion has been drawn on the performance of
DD with arbitrary timing and shaping for the general cases of
soft-cutoff noise.

In this paper,  we address the general question of the performance
of DD against soft-cutoff noise based on the general modulation function induced by arbitrary DD with bounded-strength or pulsed control. We show that for the noise spectrum with a power-law asymptote at high frequencies, there exists no modulation function to eliminate
the decoherence to an arbitrary order of the short time, regardless
of the timing and shaping of the DD control under the population
conserving condition.  Although the decohernce can be suppressed to be arbitrarily small by DD with a sufficiently large number of pulses, the existence of the largest achievable decoupling order shows that DD against soft-cutoff noise does not
have the order-by-order decoupling efficiency, which is possible for hard-cutoff noise.  Since for soft-cutoff
noise the decoherence cannot be eliminated at a certain order of short
time, we derive a set of equations to minimize the leading-order term
in the short-time expansion while eliminating the lower orders. These
equations are numerically solved for optimal solutions. In particular,
for noise correlations with a linear order term in time, we prove
that the CPMG sequences are optimal. For other noise correlations
with odd-order terms, the minimum pulse interval of the optimized
sequences is larger than in UDD sequences. This feature is important
in realistic experiments when there is a minimum pulse switching time~\cite{Khodjasteh:2011:020305}.

This paper is organized as follows: In Sec.~\ref{sec:DecoherenceControl},
we analyze the performance of DD against soft-cutoff noise, and we give the condition under which decoherence suppression
to an arbitrary order of short-time scaling is impossible. The relation between the high-frequency
cutoff and the short-time expansion of correlations is also discussed.
In Sec.~\ref{sec:ShortTimeOptimization}, we derive the equations
for sequence optimization and obtain optimal DD for noise correlations
with odd-power expansion terms. Finally, the conclusions are drawn
in Sec.~\ref{sec:Summary-and-conclusions}.

\section{No-go theorem on dynamical decoupling against noise with soft cutoff\label{sec:DecoherenceControl} }

We consider the pure-dephasing Hamiltonian for a single spin (qubit)
\begin{equation}
H=\frac{1}{2}\sigma_{z}[\omega_{a}+\beta(t)],\label{eq:HPureDephasing}
\end{equation}
 where $\sigma_{z}=|+\rangle\langle+|-|-\rangle\langle-|$ is the
Pauli operator of the qubit, $\omega_{a}$ is the energy splitting
of the qubit, and $\beta(t)$ describes random noise with average
$\overline{\beta(t)}=0$. Here the over bar denotes averaging over
the noise realizations. We assume that the statistics of the noise
fluctuations are Gaussian.

After a duration of free evolution time $T$, the noise induces between
the two states $|\pm\rangle$ a random phase shift $\int_{0}^{T}\beta(t)$
that destroys the quantum coherence. We can suppress the decoherence
by DD control on the qubit. There is only one noise source $\beta(t)$
in the model Eq.~(\ref{eq:HPureDephasing}), and to suppress the
decoherence we need to protect a MOOS which consists of a Pauli operator
$\sigma_{x}$ (more generally $\sigma_{x}\cos\phi+\sigma_{y}\sin\phi$
with $\phi$ being real)~\cite{Wang:2011:022306}. We will prove
later that DD can suppress the decoherence (i.e., the protection of
the MOOS $\{\sigma_{x}\}$) only to a certain order of short evolution
time for noise correlations that have odd-power expansion terms in
time. We expect that the proof also applies to other quantum systems
(e.g., multi-qubit systems) when the noise correlations have odd-power
terms, since in those systems there are more noise sources and more
system operators (e.g., a MOOS consisting of $L>1$ Pauli operators)
should be protected.

When we apply a sequence of instantaneous unitary operations $\sigma_{x}$
at the moments $T_{1}$, $T_{2}$, $\ldots$, $T_{N}$, the controlled
evolution operator reads
\begin{eqnarray}
U(T) & = & (\sigma_{x})^{N}U(T_{N+1},T_{N})\sigma_{x}U(T_{N},T_{N-1})\cdots\nonumber \\
 &  & \times\sigma_{x}U(T_{2},T_{1})\sigma_{x}U(T_{1},T_{0}),
\end{eqnarray}
 where $T_{0}=0$, $T_{N+1}=T$, and the free evolution operator %\begin{equation}
%U(T_{j+1},T_{j})=\exp\left\{-i\frac{\sigma_{z}}{2}\int_{T_{j}}^{T_{j+1}}\left[\omega_{a}+\beta(t)\right]dt\right\}.
%\end{equation}
\begin{equation}
U(T_{j+1},T_{j})=e^{-i\frac{\sigma_{z}}{2}\int_{T_{j}}^{T_{j+1}}\left[\omega_{a}+\beta(t)\right]dt}.
\end{equation}
 Note that when $N$ is odd, we may apply an additional $\sigma_{x}$
pulse at the end of the sequence for the identity evolution. Using
\begin{equation}
\sigma_{x}U(T_{j+1},T_{j})\sigma_{x}=e^{-i\frac{\sigma_{z}}{2}\int_{T_{j}}^{T_{j+1}}\left[-\omega_{a}-\beta(t)\right]dt},
\end{equation}
 we write the evolution operator as
\begin{equation}
U(T)=e^{-i\frac{\sigma_{z}}{2}\int_{0}^{T}\omega_{a}F_{\pi}(t/T)dt}e^{-i\frac{\sigma_{z}}{2}\int_{0}^{T}\beta(t)F_{\pi}(t/T)dt},
\end{equation}
 where we have defined the modulation function for instantaneous $\pi$-pulse
sequences~\cite{Uhrig:2007:100504,Cywinski:2008:174509}
\begin{equation}
F_{\pi}(t/T)=\begin{cases}
(-1)^{j} & \text{for }t\in(T_{j},T_{j+1}]\\
0 & \text{for }t>T,\mbox{\text{ or }}t\leq0
\end{cases}.\label{eq:Ft}
\end{equation}
The DD control is parametrized by the relative pulse
locations $T_{j}/T$. At the moment $T$, the off-diagonal density
matrix element of an ensemble is
\begin{equation}
\rho_{\uparrow\downarrow}(T)=\rho_{\uparrow\downarrow}(0)e^{-i\int_{0}^{T}\omega_{a}F_{\pi}(t/T)dt}\overline{e^{-i\int_{0}^{T}\beta(t)F_{\pi}(t/T)dt}}.\label{eq:Sec-DDQS-1qubit-rho}
\end{equation}
The coherence is characterized by the ensemble-averaged phase factor
\begin{equation}
W(T)\equiv\overline{e^{-i\int_{0}^{T}\beta(t)F_{\pi}(t/T)dt}}.
\end{equation}
For Gaussian noise, the ensemble-averaged phase factor $W(T)$ is
determined by the two-point correlation function $\overline{\beta(t_{1})\beta(t_{2})}$
and $W(T)$ becomes~\cite{Anderson:1954:316,Kubo:1954:935,Cywinski:2008:174509}
\begin{equation}
W(T)=e^{-\chi_{\pi}(T)},
\end{equation}
 where the phase correlation for instantaneous pulses
\begin{equation}
\chi_{\pi}(T)=\frac{1}{2}\int_{0}^{T}dt_{1}\int_{0}^{T}dt_{2}\overline{\beta(t_{1})\beta(t_{2})}F_{\pi}\left(\frac{t_{1}}{T}\right)F_{\pi}\left(\frac{t_{2}}{T}\right).\label{eq:XTimeDomain}
\end{equation}
can be written as the overlap between the noise power spectrum and
a filter function determined by the Fourier transform of the modulation
function~\cite{Cywinski:2008:174509}.

Under DD control, the qubit is flipped at different
moments, and the random field $\beta(t)$ is modulated by the modulation
function $F_{\pi}(t/T)$. For multilevel systems, the modulation functions
resulting from instantaneous $\pi$-pulse sequences may have values
not restricted to $\{\pm1\}$ for $t\in(0,T]$ (see Appendix~\ref{sec:Modulation-Functions-in}).
In Ref.~\cite{Uhrig:2010:045001}, it is shown that for DD composed
of specially engineered finite-duration pulses, the effective modulation
functions can take values from $\{+1,-1,0\}$ alternatively. We may
also encounter effective modulation functions which are triangle wave
functions during the time of system evolution~\cite{Wang:unpublished}.
For a more general analysis, we assume that the control conserves
the populations and the phase modulation function $F_{\pi}(t/T)$
has a general form as
\begin{align}
F\left(\frac{t}{T}\right) & =\begin{cases}
\text{a bounded function} & \text{for }t\in(0,T]\\
0 & \text{otherwise}
\end{cases},\label{eq:FtGeneral}
\end{align}
which has a finite number of discontinuities. The
more general phase correlation considered in this paper reads
\begin{subequations}
\begin{eqnarray}
\chi(T) & \equiv & \frac{1}{2}\int_{0}^{T}dt_{1}\int_{0}^{T}dt_{2}\overline{\beta(t_{1})\beta(t_{2})}F^{*}\left(\frac{t_{1}}{T}\right)F\left(\frac{t_{2}}{T}\right)\label{eq:XTimeDomainComplex}\\
 & = & \Re\int_{0}^{T}dt_{1}\int_{0}^{t_{1}}dt_{2}\overline{\beta(t_{1})\beta(t_{2})}F^{*}\left(\frac{t_{1}}{T}\right)F\left(\frac{t_{2}}{T}\right),\label{eq:XTimeDomainComplexRe}
\end{eqnarray}\label{eq:XTimeDomainComplexWhole}
\end{subequations}
where we have used $\overline{\beta(t_{1})\beta(t_{2})}=\overline{\beta(t_{2})\beta(t_{1})}$
to derive Eq.~(\ref{eq:XTimeDomainComplexRe}) {[}For quantum noise,
this may not be true. But in Eq.~(\ref{eq:XTimeDomainComplex}), $t_{1}$
and $t_{2}$ can be exchanged without changing the integration. So
the noise correlation can always be symmetrized{]}. It was shown that
under suitable approximation, Eq.~(\ref{eq:XTimeDomainComplexWhole})
with a complex $F(t/T)=e^{-i\int_{0}^{t/T}V(s)ds}$ describes
the average dephasing of a qubit under bounded-strength control with the amplitude
$V(s)$~\cite{Gordon:2008:010403,Gordon:2006:398,Gordon:2007:042310,Gordon:2007:S75}. In Ref.~\cite{Gordon:2008:010403}
based on minimization of $\chi(T)$, some optimal control fields $V(s)$
were obtained for some specific noise spectra. Therefore we analyse and minimize $\chi(T)$ given by
Eq.~(\ref{eq:XTimeDomainComplexWhole}) for DD design. For the special case of instantaneous $\pi$-pulse sequences
for the dephasing of a qubit, $F(t/T)=F_{\pi}(t/T)$ and $\chi(T)=\chi_{\pi}(T)$.
%%The aim of DD design is to minimize $\chi(T)$.

We assume the noise is stationary, i.e., of time translation symmetry,
$\overline{\beta(t_{1})\beta(t_{2})}=\overline{\beta(t_{1}-t_{2})\beta(0)}$.
Another symmetry is $\overline{\beta(t)\beta(0)}=\overline{\beta(0)\beta(t)}$.
These symmetries indicate that the noise correlation is an even function
of time, i.e.,
\begin{equation}
\overline{\beta(t)\beta(0)}\equiv C_{\text{corr}}(t)=C_{\text{corr}}(|t|).
\end{equation}
 The noise correlation can be transformed from the noise power spectrum
$S(\omega)$ as
\begin{equation}
\overline{\beta(t)\beta(0)}=\int_{-\infty}^{\infty}\frac{d\omega}{2\pi}S(\omega)e^{-i\omega t}.\label{eq:CorrSpectrum}
\end{equation}
Note that both $\overline{\beta(t)\beta(0)}$ and
$S(\omega)$ are real even functions.

The general filter function is defined as the Fourier
transform of the general modulation function,
\begin{subequations}
\begin{align}
\tilde{f}(\omega T) & \equiv\frac{1}{T}\int_{-\infty}^{\infty}F\left(\frac{t}{T}\right)e^{i\omega t}dt\\
 & =\int_{0}^{1}F(s)e^{i\omega Ts}ds,\label{eq:FtSpectrum}
\end{align}
\end{subequations}
which has the power expansion \begin{subequations}
\begin{align}
\tilde{f}(\omega T)=\sum_{m=0}^{\infty}\frac{(i\omega T)^{m}}{m!}\lambda_{m},\label{eq:fwExpansion}\\
\lambda_{m}\equiv\int_{0}^{1}F(s)s^{m}ds.\label{eq:lambdaGeneralFt}
\end{align}
 \end{subequations} Eq.~(\ref{eq:FtSpectrum}) shows that $\tilde{f}(\omega T)$
is bounded by $|\tilde{f}(\omega T)|\leq\int_{0}^{1}|F(s)|ds.$

As $F(s)$ has a finite number of discontinuities,
we use integration by parts and get
\begin{equation}
\tilde{f}(u)=\frac{1}{iu}\sum_{j}\left[\left.F(s)e^{ius}\right|_{s_{j}}^{s_{j+1}}-I_{j}(u)\right],
\end{equation}
where $s_{j}$ are the discontinuous points and $|I_{j}(u)|=|\int_{s_{j}}^{s_{j+1}}F^{\prime}(s)e^{ius}ds|\leq\int_{s_{j}}^{s_{j+1}}|F^{\prime}(s)|ds\equiv \tilde{I}_{j}$
is finite.
Therefore we have
\begin{equation}
|\tilde{f}(u)|\leq a_{\tilde{f}}/u, \label{eq:fwScaling}
\end{equation}
where the coefficient $a_{\tilde{f}}=\sum_{j}\left[|F(s_{j})|+|F(s_{j+1})|+\tilde{I}_{j}\right]$ is bounded.
%Therefore we have the scaling
%\begin{equation}
%\tilde{f}(\omega T)=\mathcal{O}\left(\frac{1}{\omega T}\right),\text{ for }\omega T\gg1.\label{eq:fwScaling}
%\end{equation}

For a sequence of $N$ instantaneous $\pi$ pulses,
$\lambda_{m}$ reads
\begin{equation}
\lambda_{m}^{(\pi)}=\sum_{j=0}^{N}\frac{(-1)^{j}}{(m+1)}\left[\left(\frac{T_{j+1}}{T}\right)^{m+1}-\left(\frac{T_{j}}{T}\right)^{m+1}\right].\label{eq:lambdaMPulses}
\end{equation}
%\textcolor{cyan}{{[}$\tilde{F}(\omega)=T\tilde{f}(\omega T)$, $\Lambda_{m+1}=(m+1)T^{m+1}\lambda_{m}$
%in the old manuscript{]}

Using Eqs.~(\ref{eq:CorrSpectrum}) and (\ref{eq:FtSpectrum}),
Eq.~(\ref{eq:XTimeDomainComplex}) can be written as the overlap
of the noise spectrum and filter function
\begin{align}
\chi(T) & =T^{2}\int_{0}^{\infty}\frac{d\omega}{\pi}S(\omega)|\tilde{f}(\omega T)|^{2}.\label{eq:SFOverlap}
\end{align}
It should be stressed that in Eq.~(\ref{eq:SFOverlap}) the filter
function $\tilde{f}(\omega T)$ is general and not limited to the
case of instantaneous pulse sequences.

\subsection{Scaling of decoupling orders}

We separate the noise spectrum into two parts by a frequency $\Omega$.
As the noise spectrum $S(\omega)$ in Eq.~(\ref{eq:SFOverlap}) induces
decoherence linearly,
\begin{equation}
\chi(T)=\chi_{[0,\Omega]}(T)+\chi_{[\Omega,\infty]}(T),
\end{equation}
where
\begin{subequations}
\begin{align}
\chi_{[0,\Omega]}(T)=T^{2}\int_{0}^{\Omega}\frac{d\omega}{\pi}S(\omega)|\tilde{f}(\omega T)|^{2},\label{eq:LowFrequencyChiT}\\
\chi_{[\Omega,\infty]}(T)=T^{2}\int_{\Omega}^{\infty}\frac{d\omega}{\pi}S(\omega)|\tilde{f}(\omega T)|^{2}.\label{eq:HighFrequencyChiT}
\end{align}
\end{subequations}
$\chi_{[0,\Omega]}(T)$ and $\chi_{[\Omega,\infty]}(T)$ account for
the effects of the low- and high-frequency noise, respectively. Both
$\chi_{[0,\Omega]}(T)$ and $\chi_{[\Omega,\infty]}(T)$ cause decoherence
 as $S(\omega)\geq0$.

\subsubsection{Effects of low-frequency noise\label{sub:Effects-of-low-frequency}}

For the spectrum $S(\omega)=\mathcal{O}(1/\omega^{P})$ with $P<1$
when $\omega\rightarrow0$, Eq.~(\ref{eq:CorrSpectrum}) gives the
noise correlation of the noise with the frequencies $\omega<\Omega$,
\begin{subequations}
\begin{align}
\overline{\beta(t)\beta(0)}_{[0,\Omega]} & =\int_{0}^{\Omega}\frac{d\omega}{\pi}S(\omega)\cos\left(\omega t\right)\\
 & =\sum_{m=0}^{\infty}C_{2m}\Omega^{2m+1}t^{2m}.
\end{align}
\end{subequations}
Here the coefficients
\begin{equation}
C_{2m}=(-1)^{2m}\int_{0}^{1}\frac{du}{\pi}S(u\Omega)\frac{u^{2m}}{(2m)!},
\end{equation}
which depend on the noise spectrum $S(\omega)$ and $\Omega$, converge at low frequencies $\omega\rightarrow0$.

Eq.~(\ref{eq:XTimeDomainComplexRe}) gives
\begin{equation}
\chi_{[0,\Omega]}(T)=\sum_{m=0}^{\infty}C_{2m}\phi_{2m}\Omega^{2m+1}T^{2m+2},
\end{equation}
where the decoherence functions
\begin{equation}
\phi_{k}\equiv\Re\int_{0}^{1}ds_{1}\int_{0}^{s_{1}}ds_{2}(s_{1}-s_{2})^{k}F^{*}\left(s_{1}\right)F\left(s_{2}\right),\label{eq:phiK-initial-form}
\end{equation}
are modified by the modulation function of DD. The even-order decoherence functions $\phi_{2k}$
control the effects of the low-frequency noise.

Therefore if the modulation function $F(t/T)$ is designed to make
\begin{equation}
\{\phi_{2m}=0\}_{m=0}^{M-1}\equiv\{\phi_{0}=\phi_{2}=\cdots=\phi_{2M-2}=0\},
\end{equation}
the decoherence from low-frequency noise is eliminated to $\chi_{[0,\Omega]}(T)=\mathcal{O}(T^{2M+2})$
(the prefactor of the scaling depends on the noise spectrum and $\Omega$).
Note that $e^{-i\int_{0}^{T}\omega_{a}F(t/T)dt}=1$ in Eq.~(\ref{eq:Sec-DDQS-1qubit-rho})
when $\phi_{0}=0$.

In Appendix~\ref{sec:Analysis-of-PhiK}, we simplify the even-order
 $\phi_{2m}$ as
\begin{align}
\phi_{2m} & =\frac{(2m)!}{2}\Re\sum_{r=0}^{2m}(-1)^{r}\frac{\lambda_{r}^{*}}{r!}\frac{\lambda_{2m-r}}{(2m-r)!}.\label{eq:phiK-even}
\end{align}
From Eq.~(\ref{eq:phiK-even}), we find that the following two sets
of equations are equivalent
\begin{equation}
\{\phi_{2m}=0\}_{m=0}^{M-1}\Leftrightarrow\{\lambda_{m}=0\}_{m=0}^{M-1}.\label{eq:GeneralDDConditions}
\end{equation}
For instantaneous $\pi$-pulse sequences, the optimal solution of
the equation set $\{\lambda_{m}=\lambda_{m}^{(\pi)}=0\}_{m=0}^{N-1}$
is
\begin{equation}
T_{j}^{\text{UDD}}=T\sin^{2}\left[\frac{\pi j}{2N+2}\right],\:(j=1,2,\ldots,N),\label{eq:UDDTiming}
\end{equation}
which is the timing of UDD sequences~\cite{Uhrig:2007:100504}. The
conditions Eqs.~(\ref{eq:GeneralDDConditions}) and (\ref{eq:lambdaGeneralFt})
are more general than the one that leads to UDD and may lead to more
general designs of optimal DD.

For the power-law spectrum $S(\omega)\approx\alpha/\omega^{P}$ with $P\geq1$
at low frequencies, $\chi_{[0,\Omega]}(T)=T^{2}\int_{0}^{\Omega}\frac{d\omega}{\pi}\frac{\alpha}{\omega^{P}}|\tilde{f}(\omega T)|^{2}$
in general diverges. For the modulation function $F_{\pi}(t/T)$,
it was shown that the divergence of the integral can be eliminated
by high-order DD sequences~\cite{Pasini:2010:012309}. For the general
modulation function $F(t/T)$ under the conditions $\{\lambda_{m}=0\}_{m=0}^{M-1}$,
using Eq.~(\ref{eq:fwExpansion}) we have $|\tilde{f}(u)|^{2}\sim\mathcal{O}(u^{2M})$
and $\chi_{[0,\Omega]}(T)=\mathcal{O}(\Omega^{2M+1-P}T^{2M+2})$ when
$M>(P-1)/2$.

When the noise with the frequency $\omega>\Omega$ is negligible (i.e.,
a hard cutoff frequency $\omega_{c}=\Omega$ in the noise spectrum
and $\chi_{[\Omega,\infty]}\approx0$), the decoherence can be suppressed
order by order, and $\chi(T)\approx\chi_{[0,\Omega]}(T)$ has the
scaling $\chi(T)\approx\mathcal{O}(T^{2M+2})$ in short-time limit.

\subsubsection{Effects of high-frequency noise\label{sub:Effects-of-high-frequency}}

Consider the noise spectrum with a power-law asymptote at high frequencies
(i.e., a soft high-frequency cutoff),

\begin{equation}
S(\omega)\approx\frac{\alpha}{\omega^{P}},\text{ for }\Omega\leq\omega\leq\Omega_{\infty},\label{eq:PowerLawSpectrum}
\end{equation}
where $\Omega_{\infty}\gg1/T$ and $\Omega\lesssim1/T$. We assume
that the decay of the noise at the frequencies $\omega>\Omega_{\infty}$
is not slower than $\alpha/\omega^{P}$ and the contribution is negligible.
We set $\Omega_{\infty}=\infty$. The high-frequency contribution
Eq.~(\ref{eq:HighFrequencyChiT}) reads
\begin{equation}
\chi_{[\Omega,\infty]}(T)\approx\chi_{P}(T)\equiv T^{P+1}\int_{\Omega T}^{\infty}\frac{du}{\pi}\frac{\alpha}{u^{P}}|\tilde{f}(u)|^{2}.\label{eq:32}
\end{equation}
Since $\tilde{f}(u)=\mathcal{O}\left(1/u\right)$ when $u\rightarrow\infty$ {[}Eq.~(\ref{eq:fwScaling}){]},
$\chi_{[\Omega,\infty]}(T)$ converges when $P>-1$.

 Using $|\tilde{f}(0)|^{2}=|\lambda_{0}|^{2}$, we get $\lim_{T\rightarrow0}\chi_{[\Omega,\infty]}(T)=0$
even though the integral $\lim_{T\rightarrow0}\int_{\Omega T}^{\infty}\frac{du}{\pi}\frac{\alpha}{u^{P}}|\tilde{f}(u)|^{2}$
may diverge when $T\rightarrow0$. We have shown in Sec.~\ref{sub:Effects-of-low-frequency}
that under the conditions $\{\lambda_{m}=0\}_{m=0}^{M-1}$ with $2M+1-P>0$,
$\int_{0}^{\Omega T}\frac{du}{\pi}\frac{\alpha}{u^{P}}|\tilde{f}(u)|^{2}=\mathcal{O}(\Omega^{2M+1-P}T^{2M+1-P})$.
Therefore when $M>(P-1)/2$, in Eq.~(\ref{eq:32}) we extend the bounds of integration to $(0,\infty)$,
\begin{align}
\chi_{[\Omega,\infty]}(T) & = C_{P}^{\text{soft}}T^{P+1} - \mathcal{O}(T^{2M+1-P})=\mathcal{O}(T^{P+1}),
\end{align}
where $C_{P}^{\text{soft}}=\int_{0}^{\infty}\frac{du}{\pi}\frac{\alpha}{u^{P}}|\tilde{f}(u)|^{2}$ is bounded when $\{\lambda_{m}=0\}_{m=0}^{M-1}$ with $2M+1-P>0$ and $P>-1$.
The scaling $\chi_{[\Omega,\infty]}(T) = \mathcal{O}(T^{P+1})$
is the largest order of decoupling that can be achieved for the noise with soft cutoff. We have the following theorem.

\begin{thm} \label{thm:SoftNoiseThmGeneral}For the noise spectrum
with a power-law asymptote $\alpha/\omega^{P}$ with $P>-1$ at high
frequencies, the largest achievable decoupling order of DD with a
general modulation function given by Eq.~(\ref{eq:FtGeneral}) is
$\chi(T)=\mathcal{O}(T^{P+1})$ in short-time limit $T\rightarrow0$.
\end{thm}

This theorem holds for arbitrary non-zero modulation function $F(t/T)$,
and it shows that one cannot suppress the decoherence to arbitrary
order when the noise has a soft cutoff in the spectrum.

For example, the $1/f$ noise and the Lorentz spectrum $\alpha/(\Omega^2+\omega^2)$ correspond to the cases of $P=1$ and $P=2$, respectively. After
eliminating the effect of low-frequency noise by high-order DD which
satisfy $\lambda_{0}=0$, we achieve the largest decoupling order
$\chi_{P=1}(T)=\mathcal{O}\left(T^{2}\right)$ and $\chi_{P=2}(T)=\mathcal{O}\left(T^{3}\right)$.

Note that  Theorem~\ref{thm:SoftNoiseThmGeneral} applies to the order of short-time scaling and it does not mean that DD can not protect the coherence to arbitrarily high precision. The decoherence
can be  suppressed further by reducing the prefactor $C_{P}^{\text{soft}}$, i.e., the the overlap of the filter function
$|\tilde{f}(\omega T)|^{2}$ and noise spectrum $S(\omega)$.  We will discuss the optimization of pulse sequences based by minimizing the overlap in Sec.~\ref{sec:ShortTimeOptimization}.

The result in Ref.~\cite{Chen:2010:052324} that DD can suppress
decoherence at most to the third order of short evolution time for
telegraphlike noise is general for noise with arbitrary statistics.
In deriving Theorem~\ref{thm:SoftNoiseThmGeneral}, we have made the assumption
that the statistics of the noise are Gaussian and the decoherence
is determined by the two-point noise correlation. It would be interesting to generalize the theorem to non-Gaussian noise.

The perturbative regime $T\lesssim1/\Omega$ is limited by the technology of experiments. For power-law noise (e.g., $1/f$ noise), $\Omega\rightarrow0$ and the conclusion applies to arbitrary duration of $T$ when $M>(P-1)/2$.

\subsection{Noise correlation expansion and high-frequency cutoff}

\begin{figure}
\includegraphics[width=0.5\columnwidth]{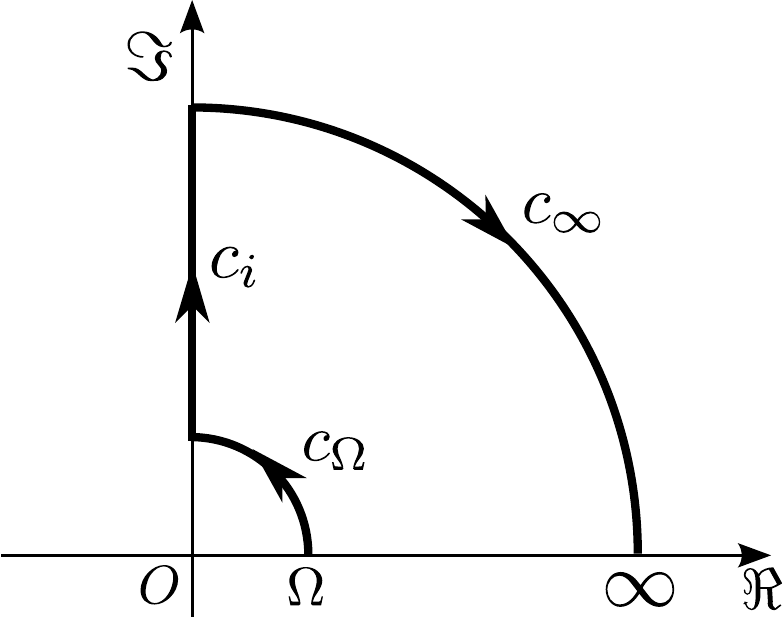}\caption{\label{fig:contours} The paths for the integral of $\alpha/z^{P}$.}
\end{figure}

The correlation in the short-time limit, which is due to high-frequency noise, can be written as
\begin{equation}
C_{\Omega,P}(t)=\Re\int_{\Omega}^{\infty}\frac{d\omega}{\pi}\frac{\alpha}{\omega^{P}}e^{-i\omega t}.
\end{equation}
 As $C_{\Omega,P}(t)$ is an even function, we just calculate the
integral for the case of $t>0$. For $P>1$ and $t>0$, we have
\begin{align*}
C_{\Omega,P}(t) & =\frac{\Re}{\pi}\left[\int_{c_{\Omega}}+\int_{c_{i}}+\int_{c_{\infty}}\right]\frac{\alpha}{z^{P}}e^{izt}dz,
\end{align*}
 where the paths $c_{\Omega}$, $c_{i}$, and $c_{\infty}$ are shown
in Fig.~(\ref{fig:contours}).

Since the maximum of $\alpha/z^{P}\rightarrow0$ as $|z|\rightarrow\infty$
in the upper half-plane, the contribution $\int_{c_{\infty}}\frac{\alpha}{z^{P}}e^{izt}dz=0$.
The contribution
\begin{equation}
\frac{\Re}{\pi}\int_{c_{i}}\frac{\alpha}{z^{P}}e^{izt}dz=\frac{\alpha}{\pi}\Re\int_{\Omega}^{\infty}i^{1-P}y^{-P}e^{-yt}dy\label{eq:CorrelationCi}
\end{equation}
 vanishes for even $P$. And
\begin{align}
\frac{\Re}{\pi}\int_{c_{\Omega}}\frac{\alpha}{z^{P}}e^{izt}dz & =\frac{\Re}{\pi}\int_{0}^{\frac{\pi}{2}}\frac{\alpha}{\Omega^{P}e^{iP\theta}}e^{i\Omega te^{i\theta}}(i\Omega)e^{i\theta}d\theta\nonumber \\
 & =I_{c_{\Omega}}^{(1)}+I_{c_{\Omega}}^{(2)},
\end{align}
 where \begin{subequations}
\begin{align}
I_{c_{\Omega}}^{(1)} & =\frac{\Re}{\pi}\sum_{r=0,r\neq P-1}^{\infty}\frac{\alpha}{r!}\frac{\Omega^{1-P}(i\Omega t)^{r}}{(r+1-P)}\left(i^{r+1-P}-1\right),\\
I_{c_{\Omega}}^{(2)} & 
=\begin{cases}
\frac{1}{2}\alpha\Re i^{P}t^{P-1}/(P-1)!, & \text{if }r \text{ is an integer,}\\
0, & \text{otherwise.}
\end{cases}
%=\frac{\alpha}{2}\Re\frac{i^{P}t^{P-1}}{(P-1)!}\sum_{r=0}^{\infty}\delta_{r,P-1}.
\end{align}
\end{subequations}
For even $P$, $I_{c_{\Omega}}^{(1)}$ is an
expansion of even powers of $t$
\begin{equation}
I_{c_{\Omega}}^{(1)}=\sum_{k=0}^{\infty} C_{2k}t^{2k}, C_{2k}=\frac{\alpha(-1)^{k+1}\Omega^{1-P+2k}}{\pi(2k)!(2k+1-P)}\neq 0,
\end{equation}
%with $C_{2k}=\frac{\alpha(-1)^{k+1}\Omega^{1-P+2k}}{\pi(2k)!(2k+1-P)}\neq 0$ and
and there is only one odd-order expansion term, which is
\begin{equation}
I_{c_{\Omega}}^{(2)}=\frac{(-1)^{P/2}\alpha}{2(P-1)!}|t|^{P-1}.\label{eq:Ic2}
\end{equation}

The existence of an odd-order term means that the correlation function
is non-analytical, which indicates that the noise source cannot be
a finite quantum bath. The noise with non-analytical correlation functions
must come from the fluctuations of an infinite bath, since otherwise
the unitary evolution of a finite quantum system will always lead
to analytical correlation functions. For example, the noise correlation
$\overline{\beta(t)\beta(0)}=e^{-|t|/t_{c}}$ has odd-order terms
in the time expansion, and the noise has a Lorentz spectrum, which
has a power-law decrease at high frequencies. This kind of noise can
be caused by Markovian (or instantaneous) collisions in the bath~\cite{Berman:1985:2784}.

As an example, we consider the following noise spectrum,
\begin{equation}
S_{2K}^{\prime}(\omega)\equiv\frac{\alpha}{\Omega_{c}^{2K}+\omega^{2K}},\label{eq:generalPowerLawNoise}
\end{equation}
 for $K\in\{1,2,\ldots\}$. $S_{2K}^{\prime}(\omega)\approx\alpha/\omega^{2K}$
when $\omega\gg\Omega_{c}$. For example, the measured ambient noise for
ions in a Penning trap has an approximate $1/\omega^{4}$ spectrum
at high frequencies and a flat dependence at low frequencies~\cite{Biercuk:2009:996},
which approximately corresponds to the noise spectrum Eq.~(\ref{eq:generalPowerLawNoise})
with $K=2$. The corresponding correlation function of Eq.~(\ref{eq:generalPowerLawNoise})
is obtained by the inverse transform
\begin{equation}
\overline{\beta(t)\beta(0)}_{2K}=\int_{0}^{\infty}\frac{d\omega}{\pi}\frac{\alpha e^{-i\omega t}}{\Omega_{c}^{2K}+\omega^{2K}}.
\end{equation}
 Using the residue theorem, we have
\begin{equation}
\overline{\beta(t)\beta(0)}_{2K}=\frac{i\alpha}{2K}\Omega_{c}^{1-2K}\sum_{n=0}^{K-1}\frac{\exp[-ie^{i\frac{\pi}{2K}(2n+1)}|\Omega_{c} t|]}{\exp[i\frac{\pi}{2K}(2n+1)(2K-1)]}.
\end{equation}
 Expanding the right-hand side in powers of $t$, we get
%\begin{multline}
%\overline{\beta(t)\beta(0)}_{2K}=\frac{i\alpha}{2K}\Omega_{c}^{1-2K}\sum_{m=0}^{\infty}\frac{(-i|\Omega_{c} t|)^{m}}{m!}\\
%\times\sum_{n=0}^{K-1}e^{i\frac{\pi}{2K}(2n+1)(m-2K+1)},
%\end{multline}
% and summing the terms gives
\begin{align}
\overline{\beta(t)\beta(0)}_{2K} & =\sum_{m=0}^{\infty}\frac{(-i|\Omega_{c} t|)^{m}}{m!}\frac{e^{i\frac{(m-2K+1)\pi}{2K}}[(-1)^{m}+1]}{e^{i(m+1)\pi/K}-1}\nonumber \\
 & \times\frac{-i\alpha}{2K}\Omega_{c}^{1-2K}.\label{eq:powerLawCoheFuncExpan}
\end{align}
 In Eq.~(\ref{eq:powerLawCoheFuncExpan}), the leading odd-order
term of the time expansion is $\frac{(-1)^{K}\alpha}{2(2K-1)!}|t|^{2K-1}$,
as predicted by Eq.~(\ref{eq:Ic2}).

For simplicity, let us consider the noise correlations
that have the power expansion

\begin{equation}
\overline{\beta(t)\beta(0)}\equiv C_{\text{corr}}(t)=\sum_{k=0}^{\infty}C_{k}\left|t\right|^{k},\label{eq:corr-expansion}
\end{equation}
where the expansion coefficients
\begin{equation}
C_{k}\equiv\left.\frac{1}{k!}\frac{d^{k}C_{\text{corr}}(t)}{dt^{k}}\right|_{t\rightarrow0^{+}}\label{eq:CkTime0}
\end{equation}
 are finite real numbers with \begin{subequations}
\begin{align}
C_{2k-1} & =0,\text{ for }k<K,\\
C_{2K-1} & \neq0.
\end{align}
\label{eq:C2kp1Vanish} \end{subequations}
The leading odd-order
term in the short-time expansion of $C_{\text{corr}}(t)$ is $C_{2K-1}|t|^{2K-1}.$
An example is $C_{\text{corr}}(t)=e^{-|t|^{3}}$ with $C_{1}=0$ and
$C_{3}=-1$. We assume that the noise correlation decreases smoothly
at long correlation times, that is,
\begin{equation}
\lim_{t\rightarrow\infty}\frac{d^{k}}{dt^{k}}C_{\text{corr}}(t)=0,\text{ for }k=0,1,\ldots,\label{eq:CkTimeInfty}
\end{equation}
 and
\begin{align}
I_{L}(\omega) & \equiv\int_{0}^{\infty}e^{i\omega t}\frac{d^{L}}{dt^{L}}C_{\text{corr}}(t)dt,\label{eq:ILIntegrate}
\end{align}
 vanishes at large $\omega$ for $L=0,1,\ldots$~\cite{Sun:1996:343}.
For the noise correlations that decay in the correlation time smoothly
without fast oscillation, $I_{L}(\omega)$ vanishes at large frequency $\omega$. For
example, the noise correlation $e^{-|t|}$ has $I_{L}(\omega)=i(-1)^{L}/(\omega+i)\rightarrow0$
when $\omega\rightarrow\infty$.

We consider the high frequency behavior of the noise spectrum, $S_{2K}(\omega)=\int_{-\infty}^{\infty}C_{\text{corr}}(t)e^{i\omega t}dt.$
Integration by parts $L\geq(2K+1)$ times gives
\begin{equation}
S_{2K}(\omega)=2\Re\left[\sum_{r=1}^{L}\frac{(r-1)!}{(-i\omega)^{r}}C_{r-1}+\frac{I_{L}(\omega)}{(-i\omega)^{L}}\right],\label{eq:j2mp1-integration-by-parts}
\end{equation}
 where we have used Eqs.~(\ref{eq:CkTime0}) and (\ref{eq:CkTimeInfty}).

Using Eqs.~(\ref{eq:C2kp1Vanish}) and (\ref{eq:ILIntegrate}), we
obtain for large $\omega$,
\begin{equation}
S_{2K}(\omega)\approx2\frac{(2K-1)!}{(i\omega)^{2K}}C_{2K-1}+\mathcal{O}\left(\frac{1}{\omega^{2K+1}}\right),
\end{equation}
 which is a power-law decrease at high frequencies.

When the noise correlation expansion contains only even-order terms,
from Eq.~(\ref{eq:j2mp1-integration-by-parts}) we have the noise
spectrum $S_{\text{even}}(\omega)=2\Re I_{L}(\omega)/(-i\omega)^{L}$
for an arbitrarily large $L$. From the assumption $\lim_{\omega\rightarrow\infty}I_{L}(\omega)=0$,
we have $\lim_{\omega\rightarrow\infty}S_{\text{even}}(\omega)\omega^{L}=0$
for an arbitrarily large $L$ and therefore the noise spectrum has
a hard high-frequency cutoff. One example is the correlation function
$e^{-t^{2}}$, which has the noise spectrum of exponential form $\sim\exp(-\frac{\omega^{2}}{4})$,
and obviously the UDD sequence can suppress the noise effect order
by order. The large $\omega$ behavior of other correlation functions
of the form $\exp(-\sum_{j=1}^{p}\alpha_{2j}t^{2j})$ can be calculated
by the saddle point integration method, which gives a result of an
exponential decrease at high frequencies (i.e., hard cutoff). For
example, when $\omega$ is very large, $\int_{-\infty}^{\infty}e^{-t^{4}}e^{i\omega t}dt\simeq\frac{1}{2}\Im\sqrt{\frac{2\pi}{a(\omega)}}e^{g(\omega)}$,
where $g(\omega)=3(\frac{\omega}{4})^{\frac{4}{3}}e^{-i2\pi/3}$ and
$a(\omega)=12(\frac{\omega}{4})^{2/3}e^{i2\pi/3}$.

\section{SEQUENCE OPTIMIZATION IN SHORT-TIME LIMIT \label{sec:ShortTimeOptimization}}

In this section, we optimize DD for the noise correlations
that have the power expansion given by Eq.~(\ref{eq:corr-expansion}),
$\overline{\beta(t)\beta(0)}\equiv\sum_{k=0}^{\infty}C_{k}\left|t\right|^{k}$.

The performance of DD in the short-time limit is directly derived
from the time-domain expansion. The time-domain expansion of noise correlations has the advantage to avoid the divergence of the decoherence $\chi(T)$.
Using the expansion Eq.~(\ref{eq:corr-expansion}),
we write
\begin{equation}
\chi(T)=\sum_{k=0}^{\infty}C_{k}\phi_{k}T^{k+2},\label{eq:chiTphiK}
\end{equation}
where the decoherence functions $\phi_{k}$ is given
by Eq.~(\ref{eq:phiK-initial-form}). It seems that we can find DD
schemes to suppress the decoherence to an arbitrary order $\chi(T)=\mathcal{O}(T^{M+2})$
by solving the equations $\phi_{k}=0$ with $C_k\neq0$ for $k< M$. However, we
have shown in Sec.~\ref{sub:Effects-of-high-frequency} by Theorem~\ref{thm:SoftNoiseThmGeneral}
that for soft-cutoff noise, there is a largest decoupling order.

The even-order functions $\phi_{2m}$ are given by
Eq.~(\ref{eq:phiK-even}). Using Eq.~(\ref{eq:phiK-general}), we
have the odd-order functions $\phi_{2M-1}$,
\begin{gather}
\phi_{2M-1}=-\frac{(2M-1)!}{2}\sum_{r=0}^{2M-1}(-1)^{r}\frac{\lambda_{r}^{*}}{r!}\frac{\lambda_{2M-1-r}}{(2M-1-r)!}\nonumber \\
+(2M-1)!\int_{-\infty}^{\infty}\frac{d\omega}{2\pi}\left[\frac{|\tilde{f}(\omega)|^{2}}{(-i\omega)^{2M}}\right.\nonumber \\
\left.-\sum_{r=0}^{2M-1}\sum_{n=0}^{2M-1-r}\frac{(-1)^{r}}{(i\omega)^{k-r-n+1}}\frac{\lambda_{r}^{*}}{r!}\frac{\lambda_{n}}{n!}\right].
\end{gather}
The condition Eq.~(\ref{eq:GeneralDDConditions}) gives
\begin{gather}
\phi_{2M-1}=(2M-1)!\int_{-\infty}^{\infty}\frac{d\omega}{2\pi}\left[\frac{|\tilde{f}(\omega)|^{2}}{(-i\omega)^{2M}}\right.\nonumber \\
\left.-\sum_{r=M}^{2M-1}\sum_{n=M}^{2M-1-r}\frac{(-1)^{r}}{(i\omega)^{k-r-n+1}}\frac{\lambda_{r}^{*}}{r!}\frac{\lambda_{n}}{n!}\right].
\end{gather}
 Notice in the summation $2M-1-r<M$. We obtain
\begin{equation}
\phi_{2M-1}=(-1)^{M}(2M-1)!\int_{-\infty}^{\infty}\frac{d\omega}{2\pi}\frac{|\tilde{f}(\omega)|^{2}}{\omega^{2M}}.\label{eq:oddPhiRemain}
\end{equation}
In Eq.~(\ref{eq:oddPhiRemain}), the integrand $|\tilde{f}(\omega)|^{2}/\omega^{2M}\geq0$
and it cannot vanish for all $\omega$ from $-\infty$ to $\infty$.
Thus we have the following theorem.

\begin{thm} \label{thm:SoftNoiseThm}There is no non-zero modulation function $F(t/T)$ to eliminate the errors so that the
equations $\{\phi_{2m}=0\}_{m=0}^{M-1}$ and $\phi_{2M-1}=0$ satisfy
simultaneously. \end{thm}

For example, for the noise with the correlation function $e^{-|t/t_{c}|}$,
one cannot simultaneously eliminate the two leading decoherence terms
$C_{0}\phi_{0}$ and $C_{1}\phi_{1}$, and the error induced by the
noise is at least $\mathcal{O}(T^{3})$. The result is
consistent with Theorem~\ref{thm:SoftNoiseThmGeneral}, since the
noise correlation $e^{-|t/t_{c}|}$ implies a spectrum with a soft
cutoff at high frequencies $S(\omega)=\frac{t_{c}}{1+(\omega t_{c})}$.

\subsection{Sequence optimization}

In this paper, we optimize the DD performance in the short-time limit. As indicated in Eq.~(\ref{eq:oddPhiRemain}),
a smaller $\tilde{f}(\omega T)$ at low frequencies will give a smaller
$\phi_{2M-1}$. Here we focus on DD with ideal instantaneous $\pi$
pulses. We use more pulses to construct a more efficient modulation
function $F(t/T)=F_{\pi}(t/T)$ to minimize $\phi_{2M-1}$, with the
conditions $\{\phi_{2m}=0\}_{m=0}^{M-1}$ {[}Eq.~(\ref{eq:GeneralDDConditions}){]}.
Using the method of Lagrange multipliers, we need to solve a set of
nonlinear equations as \begin{subequations}
\begin{gather}
\nabla_{\{y,T\}}G_{M}=0,\\
G_{M}\equiv\sum_{j=0}^{M-1}y_{j}\lambda_{j}^{(\pi)}+\phi_{2M-1},\\
\nabla_{\{y,T\}}\equiv(\frac{\partial}{\partial T_{1}},...,\frac{\partial}{\partial T_{N}},\frac{\partial}{\partial y_{0}},...,\frac{\partial}{\partial y_{M-1}}).
\end{gather}
\label{eq:shortTimeOptimization} \end{subequations} The introduced
variables $\{y_{j}\}$ are the Lagrange multipliers. Note that the
sequence optimization in Ref.~\cite{Pasini:2010:012309} also used
the constraints $\{\lambda_{m}^{(\pi)}=0\}$, but the constraints
were used there to guarantee the convergence of the calculation of
$\chi(T)$. Here, the constraints eliminate the lowest orders of errors
($\{\phi_{2m}=0\}_{m=0}^{M-1}$) {[}see Eq.~(\ref{eq:GeneralDDConditions}){]}
in short-time limit. In particular, the decoherence from inhomogeneous
broadening is eliminated when $\phi_{0}=0$.

We calculate Eq.~(\ref{eq:phiK-initial-form}) by separating the
domain of integration according to the value of $F_{\pi}(t_{1}/T)F_{\pi}(t_{2}/T)$.
For $k\geq0$, we obtain
\begin{align}
\phi_{k} & =\frac{-1}{T^{k+2}(k+1)(k+2)}\left[4\sum_{j=2}^{N}\sum_{i=1}^{j-1}(T_{j}-T_{i})^{k+2}(-1)^{i+j}\right.\nonumber \\
 & +(T_{N+1}-T_{0})^{k+2}(-1)^{N+1}+2\sum_{j=1}^{N}(T_{j}-T_{0})^{k+2}(-1)^{j}\nonumber \\
 & \left.+2\sum_{i=1}^{N}(T_{N+1}-T_{i})^{k+2}(-1)^{N+1+i}\right].\label{eq:phiKTimeDomain}
\end{align}
 Then we have
\begin{gather}
\frac{\partial\phi_{2M-1}}{\partial(T_{k}/T)}=\frac{(-1)^{k}}{T^{2M}M}\left[2\sum_{j=k+1}^{N}(T_{j}-T_{k})^{2M}(-1)^{j}-(T_{k}-T_{0})^{2M}\right.\nonumber \\
\left.+(T_{N+1}-T_{k})^{2M}(-1)^{N+1}-2\sum_{j=1}^{k-1}(T_{k}-T_{j})^{2M}(-1)^{j}\right].\label{eq:optimizeDerivative}
\end{gather}

For the special case of $M=1$, $G_{1}=y_{0}\lambda_{0}^{(\pi)}+\phi_{1}$,
we find that the CPMG sequences are solutions to Eq.~(\ref{eq:shortTimeOptimization}).
The timing of an $N$-pulse CPMG sequence reads
\begin{equation}
T_{j}^{\text{CPMG}}=\frac{2j-1}{2N}T,\text{ for }j=1,...,N.
\end{equation}
 The CPMG sequences obviously satisfy the constraint $\lambda_{0}^{(\pi)}=0$
{[}see Eq.~(\ref{eq:lambdaMPulses}){]}, which is the so-called echo
condition that eliminates the effect of static inhomogeneous broadening.
Eqs.~(\ref{eq:optimizeDerivative}) and (\ref{eq:lambdaMPulses})
give \begin{subequations}
\begin{gather}
\left.\frac{\partial\phi_{1}}{\partial(T_{k}/T)}\right|_{\text{CPMG}}=(-1)^{k+1}\frac{1}{4N^{2}}\left[1+(-1)^{N}\right],\\
\left.y_{0}\frac{\partial\lambda_{0}^{(\pi)}}{\partial(T_{k}/T)}\right|_{\text{CPMG}}=2(-1)^{k+1}y_{0}.
\end{gather}
 \end{subequations} Thus the CPMG sequences also satisfy Eq.~(\ref{eq:shortTimeOptimization})
with $y_{0}=-\frac{1}{8N^{2}}\left[1+(-1)^{N}\right]$, so they are
at least the locally optimal pulse sequences. It has been proved that
CPMG sequences are the most efficient pulse sequences in protecting
the qubit coherence against telegraph-like noise in the short-time
limit~\cite{Chen:2010:052324}. With numerical evidence, we conjecture
that it is also true that the CPMG sequences are globally optimal in the short-time
limit when the time expansion of the noise correlation function has
the two leading terms $C_{0}$ and $C_{1}|t|$.

\begin{figure}
\includegraphics[width=1\columnwidth]{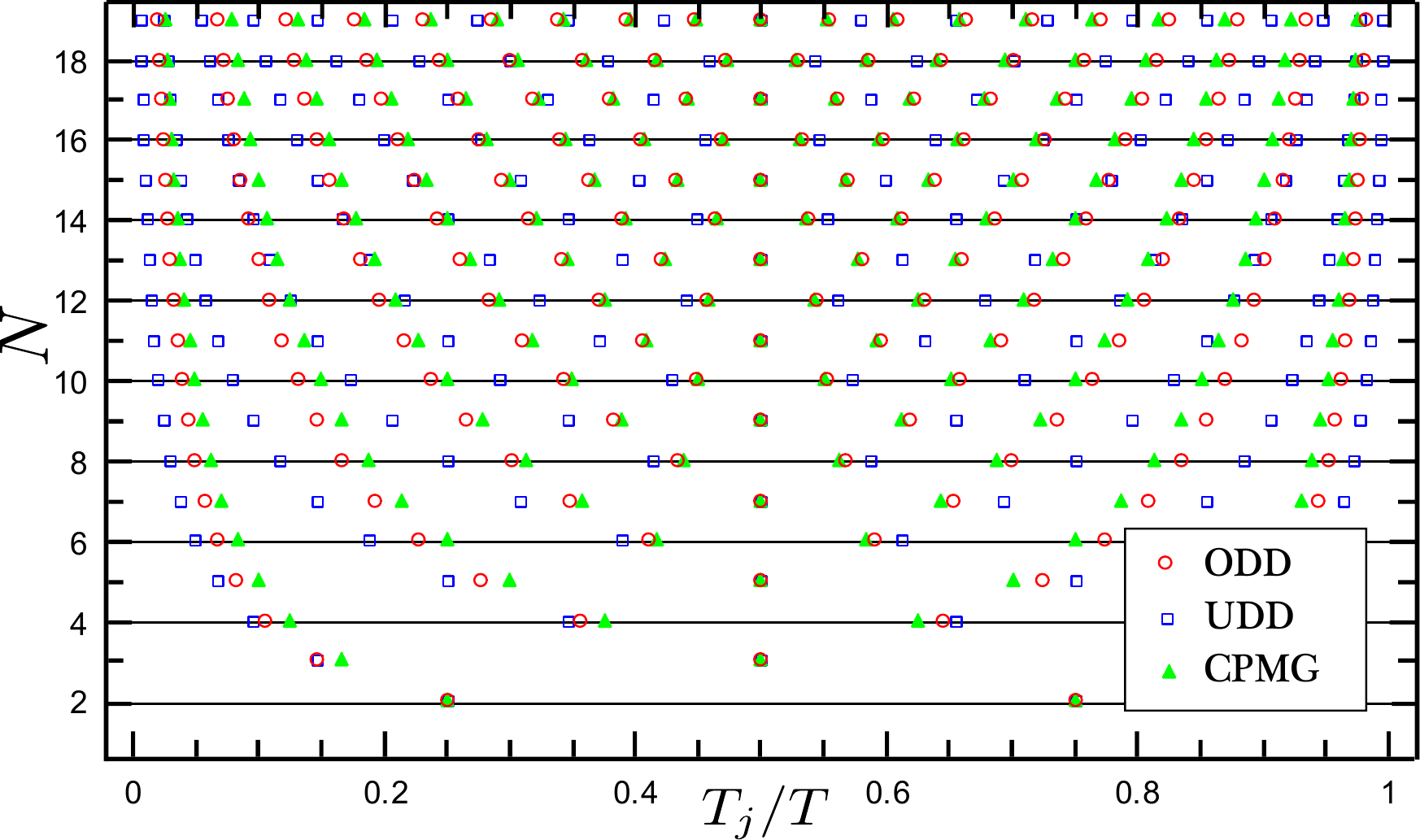}\caption{\label{fig:023RelSeqs} (Color online). Comparison of the $\text{ODD}$,
UDD and CPMG sequences for different pulse number $N$. Squares (blue),
triangles (green), and circles (red) correspond to UDD, CPMG, and
$\text{ODD}$. The $\text{ODD}$ sequences are optimized to minimize
$\phi_{3}$ under the constraints $\phi_{0}=\phi_{2}=0$.}
\end{figure}

For other cases of minimizing $\phi_{2M-1}$ with the condition $\{\lambda_{m}^{(\pi)}=0\}_{m=0}^{M-1}$,
one can see that the short-time optimized DD ($\text{ODD}$) coincides
with UDD for pulse number $N\leq M$. And as $N$ increases, the $\text{ODD}$
sequences gradually approach the CPMG sequences. For example, $\text{ODD}$
for the noise correlation
\begin{equation}
\overline{\beta(t)\beta(0)}=C_{0}+C_{2}t^{2}+C_{3}|t|^{3}+\mathcal{O}(t^{4}),\label{eq:power4noise}
\end{equation}
 is shown in Fig.~\ref{fig:023RelSeqs} in comparison with UDD and
CPMG. The $\text{ODD}$ sequences resemble the CPMG sequences when
$N$ is large.

\begin{figure}
\includegraphics[width=0.9\columnwidth]{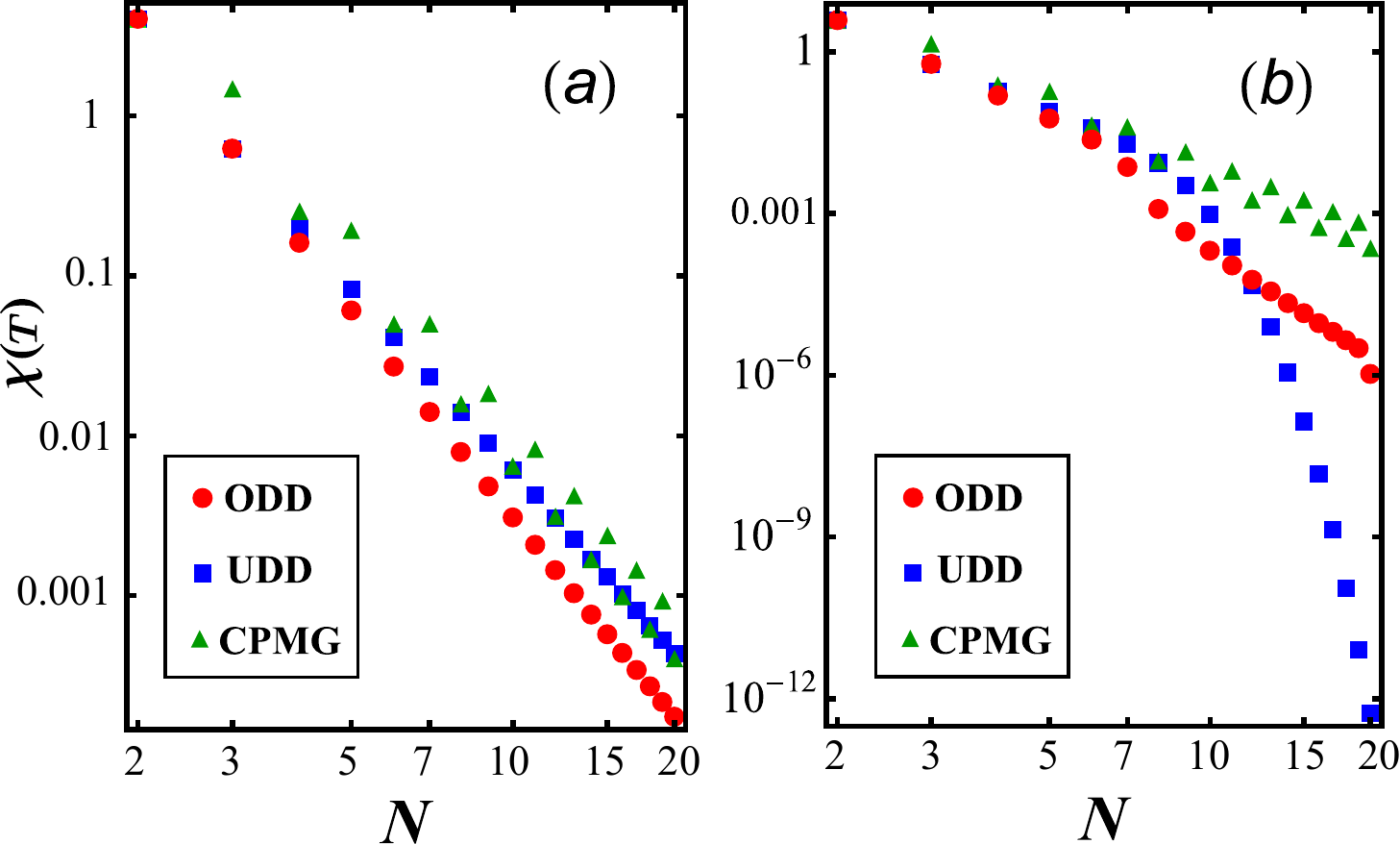}\caption{\label{fig:023eff}(Color online). The decoherence function $\chi(T)$
as a function of pulse number $N$ (a) without hard cutoff and (b)
with a hard high-frequency cutoff $\omega_{c}=40$. Here the noise
spectrum $S(\omega)=\frac{10^{5}}{1+\omega^{4}}$ and $T=0.5$. Squares
(blue), triangles (green), and circles (red) correspond to UDD, CPMG,
and $\text{ODD}$. The $\text{ODD}$ sequences are the same as those
shown in Fig.~\ref{fig:023RelSeqs}.}
\end{figure}

We show in Fig.~\ref{fig:023eff}(a) the performance of three DD
schemes against the noise described by Eq.~(\ref{eq:power4noise}).
A comparison is also shown in Fig.~\ref{fig:023eff}(b) by considering
a hard high-frequency cutoff $\omega_{c}=40$. In Fig.~\ref{fig:023eff}(a),
we can see that $\text{ODD}$ sequences give better performance than
UDD and CPMG sequences. These $\text{ODD}$ sequences are optimal
for a wide range of noise which has the noise correlation given by
Eq.~(\ref{eq:power4noise}). When we introduce a high-frequency cutoff
in the noise spectrum, as the case in Fig.~\ref{fig:023eff}(b),
the $\text{ODD}$ is the best initially when the number of pulses
$N\lesssim\omega_{c}T/2\approx10$, and the UDD sequences become better
and suppress the decoherence order by order when $N$ is large and
the hard cutoff is reached. In Fig.~\ref{fig:023eff}(b), for large
$N$ UDD is better than $\text{ODD}$, since the $\text{ODD}$ sequences
are optimized for soft-cutoff noise rather than hard-cutoff noise.
In Fig.~\ref{fig:023eff}(a) the decreasing of $\chi(T)$ is a linear
decrease in the double-logarithm plot, but in Fig.~\ref{fig:023eff}(b)
it is much faster. This confirms that DD is not so efficient against
soft-cutoff noise.

\section{Summary and conclusions\label{sec:Summary-and-conclusions}}

We have studied the dynamical decoupling control of decoherence caused
by Gaussian noise with soft cutoff in a general modulation formalism. We have proved Theorem~\ref{thm:SoftNoiseThmGeneral} which shows that, for the soft cutoff with the power-law asymptote $~\alpha/\omega^{P}$ at high frequencies, DD can only suppress decoherence to $\mathcal{O}(T^{P+1})$, where $P$ does not need to be an integer. When the short-time expansion of noise correlation
has the $(2K-1)$th odd expansion term, DD can only suppress decoherence
to $\mathcal{O}(T^{2K+1})$ (Theorem~\ref{thm:SoftNoiseThm}). The existence of odd-order terms in the
short-time expansion corresponds to a soft high-frequency cutoff ($\sim\alpha/\omega^{2K}$) in
the noise spectrum.
For these noise spectra, we have derived the equations for pulse sequence
optimization, which minimizes the leading odd-order decoherence function
and eliminates even-order decoherence functions of lower orders. The
$\text{ODD}$ sequences obtained by this method coincide with the
UDD sequences when the pulse number $N$ is small, and they resemble
CPMG sequences when $N$ is large. For the special case that the short-time
correlation function expansion has a linear term in time (i.e., a soft cutoff $\sim\alpha/\omega^{2}$), the $\text{ODD}$
sequences are exactly the CPMG sequences.

Although we derived the theorems from a pure dephasing
model, we expect that the results of the existence of the largest decoupling
order in short-time limit can be extended to the general decoherence
model (including both dephasing and relaxation) of quantum systems.
It is desirable to study the DD in suppressing the general decoherence
of quantum systems in the soft-cutoff noise in the future.
\begin{acknowledgments}
We thank Yi-Fan Luo and Bobo Wei for discussions. This work was supported
by the Hong Kong GRF CUHK402209, the CUHK Focused Investments Scheme,
and the National Natural Science Foundation of China Project No. 11028510.
\end{acknowledgments}

\appendix

\section{Modulation Functions in Multiqubit Systems}\label{sec:Modulation-Functions-in}

Consider an $L$-qubit (or $2^{L}$-level) system suffering pure dephasing
described by the Hamiltonian

\begin{equation}
H=\sum_{m=0}^{2^{L}-1}|m\rangle\langle m|[\omega_{m}+\beta_{m}(t)],
\end{equation}
 where $\omega_{m}$ is the energy and $\beta_{m}(t)$ is the fluctuation
on the state $|m\rangle$. Here $m=(m_{L}\cdots m_{2}m_{1})$ is a
binary code with $m_{l}=0$ or $1$ for the $l$th qubit.

The Pauli operator for the $l$th qubit is
\begin{equation}
\sigma_{x}^{(l)}=\sum_{m_{l}=0}\left(|m+2^{l-1}\rangle\langle m|+\text{H.c.}\right),
\end{equation}
 which exchanges two basis states $|m\rangle$ and $|m^{\prime}\rangle$
if $m$ and $m^{\prime}$ differ at and only at the $l$th bit.

After a sequence of $\sigma_{x}^{(l)}$ pulses and a final pulse $\sigma_{\text{add}}=\sigma_{x}^{(l_{1})}\cdots\sigma_{x}^{(l_{N})}$,
the evolution operator is
\begin{align}
U(T) & =\sigma_{\text{add}}U(T_{N+1},T_{N})\sigma_{x}^{(l_{N})}U(T_{N},T_{N-1})\cdots\nonumber \\
 & \times\sigma_{x}^{(l_{2})}U(T_{2},T_{1})\sigma_{x}^{(l_{1})}U(T_{1},T_{0})\sigma_{x}^{(l_{0})},
\end{align}
 where $T_{0}=0$, $T_{N+1}\equiv T$, and $\sigma_{x}^{(l_{0})}\equiv I$.
The free evolution operator
\begin{equation}
U(T_{j+1},T_{j})=\exp\left[-i\int_{T_{j}}^{T_{j+1}}\sum_{m=0}^{2^{L}-1}|m\rangle\langle m|[\omega_{m}+\beta_{m}(t)]dt\right].
\end{equation}
 We write the evolution operator as
\begin{equation}
U(T)=e^{-i\sum_{j=0}^{N}\int_{T_{j}}^{T_{j+1}}\sum_{m=0}^{2^{L}-1}\Xi_{m,j}[\omega_{m}+\beta_{m}(t)]dt},
\end{equation}
with $\Xi_{m,j}\equiv\sigma_{x}^{(l_{0})}\cdots\sigma_{x}^{(l_{j})}|m\rangle\langle m|\sigma_{x}^{(l_{j})}\cdots\sigma_{x}^{(l_{0})}$.
The phase factor between the states $|p\rangle$ and $|q\rangle$
changed during the evolution time $T$ is
\begin{equation}
\varphi_{pq}(T)=\langle p|U(T)|p\rangle\langle q|U^{\dagger}(T)|q\rangle,
\end{equation}
 where
\begin{equation}
\langle p|U(T)|p\rangle=e^{-i\sum_{j=0}^{N}\int_{T_{j}}^{T_{j+1}}\langle p|\sum_{m=0}^{2^{L}-1}\Xi_{m,j}|p\rangle[\omega_{m}+\beta_{m}(t)]dt}.
\end{equation}
 Note that $\sigma_{x}^{(l_{j})}\cdots\sigma_{x}^{(l_{0})}|p\rangle=|p\oplus[l_{0}\cdots l_{j}]\rangle$
with
\begin{equation}
p\oplus[l_{0}\cdots l_{j}]\equiv p\oplus2^{l_{1}}\oplus2^{l_{2}}\cdots\oplus2^{l_{j}},
\end{equation}
for $j>0$ and $p\oplus[l_{0}]\equiv p$. Here $\oplus$ denotes addition
on binary digits, that is, $p\oplus[l_{0}\cdots l_{j}]$ is obtained
by flipping the $l_{1}$th,$\ldots$,$l_{j}$th bits of $p=(p_{L}\cdots p_{2}p_{1})$
in binary code. We obtain
\begin{equation}
\langle p|U(T)|p\rangle=e^{-i\sum_{j=0}^{N}\int_{T_{j}}^{T_{j+1}}[\omega_{p\oplus[l_{0}\cdots l_{j}]}+\beta_{p\oplus[l_{0}\cdots l_{j}]}(t)]dt}.
\end{equation}
 Therefore the coherence between the states $|p\rangle$ and $|q\rangle$
decreases by the average of the random phase
\begin{equation}
\overline{e^{-i\sum_{j=0}^{N}\int_{T_{j}}^{T_{j+1}}[\beta_{p\oplus[l_{0}\cdots l_{j}]}(t)-\beta_{q\oplus[l_{0}\cdots l_{j}]}(t)]dt}}.
\end{equation}

When each of the qubits feels the same noise, $\beta_{m}(t)=(\sum_{k=1}^{L}m_{k})\beta(t)$
for the $m$th level, and
\begin{equation}
\overline{\varphi_{pq}(T)}=e^{-i\int_{0}^{T}F_{pq}(t)\omega dt}\overline{e^{-i\int_{0}^{T}F_{pq}(t)\beta(t)dt}},
\end{equation}
where the the modulation function is defined as
\begin{equation}
F_{pq}(t)=(\sum_{k=1}^{L}\tilde{p}_{k})-(\sum_{k=1}^{L}\tilde{q}_{k}),
\end{equation}
with $\tilde{p}=p\oplus[l_{0}\cdots l_{j}]$ and $\tilde{q}=q\oplus[l_{0}\cdots l_{j}]$
for $t\in(T_{j},T_{j+1}]$. For example, when $L=2$, $F_{pq}(t)\in\{0,\pm1,\pm2\}$.

\section{Decoherence Functions $\phi_{k}$\label{sec:Analysis-of-PhiK}}

As $F(t/T)=0$ for $t\notin(0,T]$, we extend the bounds of integration
for $t$ to infinity and transform Eq.~(\ref{eq:phiK-initial-form})
to
\begin{eqnarray}
\phi_{k} & = & \Re\frac{\partial^{k}}{\partial(i\eta)^{k}}\int_{-\infty}^{\infty}\frac{d\omega_{1}}{2\pi}\int_{-\infty}^{\infty}\frac{d\omega_{2}}{2\pi}\int_{-\infty}^{\infty}dt_{1}\int_{0}^{t_{1}}dt_{2}\nonumber \\
 &  & \times\tilde{f}^{*}(\omega_{1})\tilde{f}(\omega_{2})e^{i\omega_{1}t_{1}}e^{-i\omega_{2}t_{2}}e^{i\eta(t_{1}-t_{2})},
\end{eqnarray}
 where we set $\eta\rightarrow0$ after differentiation. Integrations
over $t_{2}$, $t_{1}$ and $\omega_{1}$ give
\begin{equation}
\phi_{k}=\Re\frac{\partial^{k}}{\partial(i\eta)^{k}}\int_{-\infty}^{\infty}\frac{d\omega}{2\pi}\frac{\tilde{f}(\omega)}{i(\omega+\eta)}[\tilde{f}^{*}(-\eta)-\tilde{f}^{*}(\omega)].\label{eq:phiK_Eta}
\end{equation}
Using the formulas 
\begin{equation}
\frac{\partial^{k}}{\partial(i\eta)^{k}}\left[u(\eta)v(\eta)\right]=\sum_{r=0}^{k}\frac{k!}{r!(k-r)!}\left[\frac{\partial^{k-r}u(\eta)}{\partial(i\eta)^{k-r}}\right]\left[\frac{\partial^{r}v(\eta)}{\partial(i\eta)^{r}}\right],
\end{equation}
\begin{equation}
\left.\frac{\partial^{k-r}}{\partial(i\eta)^{k-r}}\frac{1}{i(\omega+\eta)}\right|_{\eta=0}=\frac{-(k-r)!}{(-i\omega)^{k-r+1}},\text{ for }r\geq0,
\end{equation}
and
\begin{equation}
\left.\frac{\partial^{r}}{\partial(i\eta)^{r}}\tilde{f}^{*}(-\eta)\right|_{\eta=0}=\lambda_{r}^{*},\text{ for }r\geq0,
\end{equation}
we have
\begin{equation}
\phi_{k}=k!\Re\int_{-\infty}^{\infty}\frac{d\omega}{2\pi}[\frac{|\tilde{f}(\omega)|^{2}}{(-i\omega)^{k+1}}-\sum_{r=0}^{k}\frac{\tilde{f}(\omega)}{(-i\omega)^{k-r+1}}\frac{\lambda_{r}^{*}}{r!}].
\end{equation}
Changing the summation index, we obtain
\begin{equation}
\phi_{k}=k!\Re\int_{-\infty}^{\infty}\frac{d\omega}{2\pi}\left[\frac{|\tilde{f}(\omega)|^{2}}{(-i\omega)^{k+1}}-\sum_{r=0}^{k}\frac{\tilde{f}(\omega)}{(-i\omega)^{k-r+1}}\frac{\lambda_{r}^{*}}{r!}\right].\label{eq:phiK-generalUnsimplified}
\end{equation}
Note that the summation over $r$ and the integration over frequency
cannot be exchanged when the integration does not converge for each
individual term. Using Eq.~(\ref{eq:fwExpansion}), we have
\begin{eqnarray}
\phi_{k} & = & k!\Re\int_{-\infty}^{\infty}\frac{d\omega}{2\pi}\left[\frac{|\tilde{f}(\omega)|^{2}}{(-i\omega)^{k+1}}+\sum_{r=0}^{k}\sum_{n=0}^{k-r}\frac{(-1)^{k-r}}{(i\omega)^{k-r-n+1}}\frac{\lambda_{r}^{*}}{r!}\frac{\lambda_{n}}{n!}\right.\nonumber \\
 & + & \left.\sum_{r=0}^{k}\sum_{n=k-r+1}^{\infty}\frac{(-1)^{k-r}}{(i\omega)^{k-r-n+1}}\frac{\lambda_{r}^{*}}{r!}\frac{\lambda_{n}}{n!}\right].\label{eq:phiK-general-a}
\end{eqnarray}
 We simplify the last line by using Eq.~(\ref{eq:lambdaGeneralFt})
and the equality
\begin{equation}
\int_{-\infty}^{\infty}\frac{d\omega}{2\pi}\sum_{n=r}^{\infty}\frac{(\pm i\omega)^{n-r}}{n!}t^{n}=\frac{1}{2}\frac{t^{r-1}}{(r-1)!},\text{ for }r\geq1,t\geq0,\label{eq:equalitySumWT}
\end{equation}
 which is proved in Appendix~\ref{sec:Appendix_CutOffOneEquation}.
We obtain
\begin{gather}
\phi_{k}=k!\Re\int_{-\infty}^{\infty}\frac{d\omega}{2\pi}\left[\frac{|\tilde{f}(\omega)|^{2}}{(-i\omega)^{k+1}}+\sum_{r=0}^{k}\sum_{n=0}^{k-r}\frac{(-1)^{k-r}}{(i\omega)^{k-r-n+1}}\frac{\lambda_{r}^{*}}{r!}\frac{\lambda_{n}}{n!}\right]\nonumber \\
+\frac{k!}{2}\Re\sum_{r=0}^{k}(-1)^{k-r}\frac{\lambda_{r}^{*}}{r!}\frac{\lambda_{k-r}}{(k-r)!}.\label{eq:phiK-general}
\end{gather}

For even number $2m$, $\frac{|\tilde{f}(\omega)|^{2}}{(-i\omega)^{2m+1}}$
is an odd function and its integral vanishes. Eq.~(\ref{eq:phiK-general})
gives
\begin{gather}
\phi_{2m}=(2m)!\Re\int_{-\infty}^{\infty}\frac{d\omega}{2\pi}\left[\sum_{r=0}^{2m}\sum_{n=0}^{2m-r}\frac{(-1)^{r}}{(i\omega)^{2m-r-n+1}}\frac{\lambda_{r}^{*}}{r!}\frac{\lambda_{n}}{n!}\right]\nonumber \\
+\frac{(2m)!}{2}\Re\sum_{r=0}^{2m}(-1)^{r}\frac{\lambda_{r}^{*}}{r!}\frac{\lambda_{2m-r}}{(2m-r)!},
\end{gather}
 which is decomposed as (with the changes of summation order and indices)
\begin{eqnarray}
\phi_{2m} & = & (2m)!\Re\int_{-\infty}^{\infty}\frac{d\omega}{2\pi}\left[\sum_{r=0}^{2m}\sum_{n=0}^{2m-r}\frac{(-1)^{r}/2}{(i\omega)^{2m-r-n+1}}\frac{\lambda_{r}^{*}}{r!}\frac{\lambda_{n}}{n!}\right.\nonumber \\
 &  & +\left.\sum_{r=0}^{2m}\sum_{n=0}^{2m-r}\frac{(-1)^{n}/2}{(i\omega)^{2m-r-n+1}}\frac{\lambda_{r}}{r!}\frac{\lambda_{n}^{*}}{n!}\right]\nonumber \\
 &  & +\frac{(2m)!}{2}\Re\sum_{r=0}^{2m}(-1)^{r}\frac{\lambda_{r}^{*}}{r!}\frac{\lambda_{2m-r}}{(2m-r)!},.\label{eq:sepIntPhiAs2}
\end{eqnarray}
 The integrals of the terms with odd-power of $\omega$ vanish. And
for even functions of $\omega$, the sum $n+r$ is an odd number,
so $(-1)^{r}+(-1)^{n}=0$. Thusthe real part of the
integral in Eq.~(\ref{eq:sepIntPhiAs2}) vanishes. From the last
line of Eq.~(\ref{eq:sepIntPhiAs2}), we obtain Eq.~(\ref{eq:phiK-even}),
i.e.,
\begin{align}
\phi_{2m} & =\frac{(2m)!}{2}\sum_{r=0}^{2m}(-1)^{r}\frac{\lambda_{r}^{*}}{r!}\frac{\lambda_{2m-r}}{(2m-r)!}.
\end{align}

\section{Proof of Equation (\ref{eq:equalitySumWT})\label{sec:Appendix_CutOffOneEquation}}

To prove Eq.~(\ref{eq:equalitySumWT}), we just need to prove %\begin{equation}
%\lim_{R\rightarrow\infty}\int_{-R}^{R}d\omega\sum_{n=r}^{\infty}\frac{(\pm i\omega)^{n-r}}{n!}t^{n}=\frac{\pi}{(r-1)!}t^{r-1},\text{ for }r\geq1,
%\end{equation}
\begin{equation}
\lim_{R\rightarrow\infty}\int_{-R}^{R}dx\sum_{n=r}^{\infty}\frac{(\pm ix)^{n-r}}{n!}=\frac{\pi}{(r-1)!},\text{ for }r\geq1,
\end{equation}
 where the bounds in the integral guarantee that the modulation function
$F(t)$ is a real function. Using %\begin{equation}
%\int_{-R}^{R}d\omega\sum_{n=r}^{\infty}\frac{(\pm i\omega)^{n-r}}{n!}t^{n}=\sum_{n=1}^{\infty}\frac{(Rt)^{n}t^{r-1}}{(n+r-1)!n}(i^{n-1}+\text{c.c.}),
%\end{equation}
%\begin{equation}
%\frac{t^{r-1}}{(r-1)!}\int_{-Rt}^{Rt}\frac{\sin x}{x}dx=\sum_{n=1}^{\infty}\frac{(Rt)^{n}}{n!n}\frac{t^{r-1}}{(r-1)!}(i^{n-1}+\text{c.c.}),
%\end{equation}
\begin{subequations}
\begin{gather}
\int_{-R}^{R}dx\sum_{n=r}^{\infty}\frac{(\pm ix)^{n-r}}{n!}=\sum_{n=1}^{\infty}\frac{R^{n}}{(n+r-1)!n}(i^{n-1}+\text{c.c.}),\\
\frac{1}{(r-1)!}\int_{-R}^{R}\frac{\sin x}{x}dx=\sum_{n=1}^{\infty}\frac{R^{n}}{n!n}\frac{1}{(r-1)!}(i^{n-1}+\text{c.c.}),
\end{gather}
 \end{subequations} and $\lim_{R\rightarrow\infty}\int_{-R}^{R}\frac{\sin x}{x}dx=\pi$,
we just need to prove
\begin{equation}
\lim_{R\rightarrow\infty}\sum_{n=1}^{\infty}\left[\frac{R^{n}}{(n+r-1)!n}-\frac{R^{n}}{n!n(r-1)!}\right](i^{n-1}+\text{c.c.})=0.
\end{equation}
 For $r=1$, it obviously holds. For $r\geq2$, we can show the difference
\begin{equation}
\Delta\equiv\sum_{n=1}^{\infty}\frac{R^{n}(i^{n-1}+\text{c.c.})}{(n+r-1)!n}\left[(r-1)!-\prod_{j=1}^{r-1}(n+j)\right]=\mathcal{O}\left(\frac{1}{R}\right),\label{eq:proofIdentityGoal}
\end{equation}
 so $\text{lim}_{R\rightarrow\infty}\Delta=0$. By expanding the terms
in the square brackets of Eq.~(\ref{eq:proofIdentityGoal}), we have
\begin{equation}
\Delta=\sum_{n=1}^{\infty}\frac{R^{n}(i^{n-1}+\text{c.c.})}{(n+r-1)!}\left(\sum_{k=0}^{r-2}a_{k}n^{k}\right),
\end{equation}
 where $a_{k}$ is a number independent of $n$. We arrange the terms
in the square brackets and get
\begin{equation}
\Delta=\sum_{n=1}^{\infty}\frac{R^{n}(i^{n-1}+\text{c.c.})}{(n+r-1)!}\left[\sum_{k=1}^{r-2}b_{k}\prod_{j=1}^{k}[(n+r-j)]+b_{0}\right],
\end{equation}
 with $b_{j}$ independent of $n$. After some simplification it becomes
for $r\geq2$
\begin{equation}
\sum_{k=1}^{r-1}\frac{b_{r-k-1}}{i^{k+1}R^{k}}\left(e^{iR}-\sum_{n=0}^{k-1}\frac{R^{n}}{n!}i^{n}\right)+\text{c.c.}=\mathcal{O}\left(\frac{1}{R}\right).
\end{equation}
 Hence $\Delta=\mathcal{O}(\frac{1}{R})$, and Eq.~(\ref{eq:equalitySumWT})
is proved.

%\bibliographystyle{ieeetr}
%\bibliography{D2bib}

\expandafter\ifx\csname natexlab\endcsname\relax\global\long\def\natexlab#1{#1}
\fi \expandafter\ifx\csname bibnamefont\endcsname\relax \global\long\def\bibnamefont#1{#1}
\fi \expandafter\ifx\csname bibfnamefont\endcsname\relax \global\long\def\bibfnamefont#1{#1}
\fi \expandafter\ifx\csname citenamefont\endcsname\relax \global\long\def\citenamefont#1{#1}
\fi \expandafter\ifx\csname url\endcsname\relax \global\long\def\url#1{\texttt{#1}}
\fi \expandafter\ifx\csname urlprefix\endcsname\relax\global\long\def\urlprefix{URL }
\fi \providecommand{\bibinfo}[2]{#2} \providecommand{\eprint}[2][]{\url{#2}}

%\bibliographystyle{plain}
%\bibliography{D2bib}

\begin{thebibliography}{62}
\expandafter\ifx\csname natexlab\endcsname\relax\def\natexlab#1{#1}\fi
\expandafter\ifx\csname bibnamefont\endcsname\relax
  \def\bibnamefont#1{#1}\fi
\expandafter\ifx\csname bibfnamefont\endcsname\relax
  \def\bibfnamefont#1{#1}\fi
\expandafter\ifx\csname citenamefont\endcsname\relax
  \def\citenamefont#1{#1}\fi
\expandafter\ifx\csname url\endcsname\relax
  \def\url#1{\texttt{#1}}\fi
\expandafter\ifx\csname urlprefix\endcsname\relax\def\urlprefix{URL }\fi
\providecommand{\bibinfo}[2]{#2}
\providecommand{\eprint}[2][]{\url{#2}}

\bibitem[{\citenamefont{Nielsen and Chuang}(2000)}]{Nielsen:2000:Book}
\bibinfo{author}{\bibfnamefont{M.~A.} \bibnamefont{Nielsen}} \bibnamefont{and}
  \bibinfo{author}{\bibfnamefont{I.~L.} \bibnamefont{Chuang}},
  \emph{\bibinfo{title}{Quantum Computation and Quantum Information}}
  (\bibinfo{publisher}{Cambridge University, Cambridge}, \bibinfo{year}{2000}).

\bibitem[{\citenamefont{Duan and Guo}(1997)}]{Duan:1997:DFS}
\bibinfo{author}{\bibfnamefont{L.-M.} \bibnamefont{Duan}} \bibnamefont{and}
  \bibinfo{author}{\bibfnamefont{G.-C.} \bibnamefont{Guo}},
  \bibinfo{journal}{Phys. Rev. Lett.} \textbf{\bibinfo{volume}{79}},
  \bibinfo{pages}{1953} (\bibinfo{year}{1997}).

\bibitem[{\citenamefont{Zanardi}(1998)}]{Zanardi:1998:DFS}
\bibinfo{author}{\bibfnamefont{P.}~\bibnamefont{Zanardi}},
  \bibinfo{journal}{Phys. Rev. A} \textbf{\bibinfo{volume}{57}},
  \bibinfo{pages}{3276} (\bibinfo{year}{1998}).

\bibitem[{\citenamefont{Lidar et~al.}(1998)\citenamefont{Lidar, Chuang, and
  Whaley}}]{Lidar:1998:DFS}
\bibinfo{author}{\bibfnamefont{D.~A.} \bibnamefont{Lidar}},
  \bibinfo{author}{\bibfnamefont{I.~L.} \bibnamefont{Chuang}},
  \bibnamefont{and} \bibinfo{author}{\bibfnamefont{K.~B.}
  \bibnamefont{Whaley}}, \bibinfo{journal}{Phys. Rev. Lett.}
  \textbf{\bibinfo{volume}{81}}, \bibinfo{pages}{2594} (\bibinfo{year}{1998}).

\bibitem[{\citenamefont{Shor}(1995)}]{Shor:1995:QEC}
\bibinfo{author}{\bibfnamefont{P.~W.} \bibnamefont{Shor}},
  \bibinfo{journal}{Phys. Rev. A} \textbf{\bibinfo{volume}{52}},
  \bibinfo{pages}{2493} (\bibinfo{year}{1995}).

\bibitem[{\citenamefont{Steane}(1996)}]{Steane:1996:QEC}
\bibinfo{author}{\bibfnamefont{A.~M.} \bibnamefont{Steane}},
  \bibinfo{journal}{Proc. R. Soc. Lond. A} \textbf{\bibinfo{volume}{452}},
  \bibinfo{pages}{2551} (\bibinfo{year}{1996}).

\bibitem[{\citenamefont{Viola and Lloyd}(1998)}]{Viola:1998:PRA}
\bibinfo{author}{\bibfnamefont{L.}~\bibnamefont{Viola}} \bibnamefont{and}
  \bibinfo{author}{\bibfnamefont{S.}~\bibnamefont{Lloyd}},
  \bibinfo{journal}{Phys. Rev. A} \textbf{\bibinfo{volume}{58}},
  \bibinfo{pages}{2733} (\bibinfo{year}{1998}).

\bibitem[{\citenamefont{Ban}(1998)}]{Ban:1998:DD}
\bibinfo{author}{\bibfnamefont{M.}~\bibnamefont{Ban}}, \bibinfo{journal}{J.
  Mod. Opt.} \textbf{\bibinfo{volume}{45}}, \bibinfo{pages}{2315}
  (\bibinfo{year}{1998}).

\bibitem[{\citenamefont{Zanardi}(1999)}]{Zanardi:1999:77}
\bibinfo{author}{\bibfnamefont{P.}~\bibnamefont{Zanardi}},
  \bibinfo{journal}{Phys. Lett. A} \textbf{\bibinfo{volume}{258}},
  \bibinfo{pages}{77} (\bibinfo{year}{1999}).

\bibitem[{\citenamefont{Viola et~al.}(1999)\citenamefont{Viola, Knill, and
  Lloyd}}]{Viola:1999:2417}
\bibinfo{author}{\bibfnamefont{L.}~\bibnamefont{Viola}},
  \bibinfo{author}{\bibfnamefont{E.}~\bibnamefont{Knill}}, \bibnamefont{and}
  \bibinfo{author}{\bibfnamefont{S.}~\bibnamefont{Lloyd}},
  \bibinfo{journal}{Phys. Rev. Lett.} \textbf{\bibinfo{volume}{82}},
  \bibinfo{pages}{2417} (\bibinfo{year}{1999}).

\bibitem[{\citenamefont{Yang et~al.}(2011)\citenamefont{Yang, Wang, and
  Liu}}]{Yang:2010:2}
\bibinfo{author}{\bibfnamefont{W.}~\bibnamefont{Yang}},
  \bibinfo{author}{\bibfnamefont{Z.-Y.} \bibnamefont{Wang}}, \bibnamefont{and}
  \bibinfo{author}{\bibfnamefont{R.-B.} \bibnamefont{Liu}},
  \bibinfo{journal}{Front. Phys.} \textbf{\bibinfo{volume}{6}},
  \bibinfo{pages}{2} (\bibinfo{year}{2011}).

\bibitem[{\citenamefont{Hahn}(1950)}]{Hahn:1950:Echo}
\bibinfo{author}{\bibfnamefont{E.~L.} \bibnamefont{Hahn}},
  \bibinfo{journal}{Phys. Rev.} \textbf{\bibinfo{volume}{80}},
  \bibinfo{pages}{580} (\bibinfo{year}{1950}).

\bibitem[{\citenamefont{Carr and Purcell}(1954)}]{Carr:1954:630}
\bibinfo{author}{\bibfnamefont{H.~Y.} \bibnamefont{Carr}} \bibnamefont{and}
  \bibinfo{author}{\bibfnamefont{E.~M.} \bibnamefont{Purcell}},
  \bibinfo{journal}{Phys. Rev.} \textbf{\bibinfo{volume}{94}},
  \bibinfo{pages}{630} (\bibinfo{year}{1954}).

\bibitem[{\citenamefont{Meiboom and Gill}(1958)}]{Meiboom:1958:688}
\bibinfo{author}{\bibfnamefont{S.}~\bibnamefont{Meiboom}} \bibnamefont{and}
  \bibinfo{author}{\bibfnamefont{D.}~\bibnamefont{Gill}},
  \bibinfo{journal}{Rev. Sci. Instrum.} \textbf{\bibinfo{volume}{29}},
  \bibinfo{pages}{688} (\bibinfo{year}{1958}).

\bibitem[{\citenamefont{Mehring}(1983)}]{Mehring:1983:BookNMR}
\bibinfo{author}{\bibfnamefont{M.}~\bibnamefont{Mehring}},
  \emph{\bibinfo{title}{Principles of High Resolution NMR in Solids}}
  (\bibinfo{publisher}{Spinger-Verleg}, \bibinfo{address}{Berlin},
  \bibinfo{year}{1983}), \bibinfo{edition}{2nd} ed.

\bibitem[{\citenamefont{Khodjasteh and Lidar}(2005)}]{Khodjasteh:2005:180501}
\bibinfo{author}{\bibfnamefont{K.}~\bibnamefont{Khodjasteh}} \bibnamefont{and}
  \bibinfo{author}{\bibfnamefont{D.~A.} \bibnamefont{Lidar}},
  \bibinfo{journal}{Phys. Rev. Lett.} \textbf{\bibinfo{volume}{95}},
  \bibinfo{pages}{180501} (\bibinfo{year}{2005}).

\bibitem[{\citenamefont{Khodjasteh and Lidar}(2007)}]{Khodjasteh:2007:062310}
\bibinfo{author}{\bibfnamefont{K.}~\bibnamefont{Khodjasteh}} \bibnamefont{and}
  \bibinfo{author}{\bibfnamefont{D.~A.} \bibnamefont{Lidar}},
  \bibinfo{journal}{Phys. Rev. A} \textbf{\bibinfo{volume}{75}},
  \bibinfo{pages}{062310} (\bibinfo{year}{2007}).

\bibitem[{\citenamefont{Yao et~al.}(2007)\citenamefont{Yao, Liu, and
  Sham}}]{Yao:2007:077602}
\bibinfo{author}{\bibfnamefont{W.}~\bibnamefont{Yao}},
  \bibinfo{author}{\bibfnamefont{R.-B.} \bibnamefont{Liu}}, \bibnamefont{and}
  \bibinfo{author}{\bibfnamefont{L.~J.} \bibnamefont{Sham}},
  \bibinfo{journal}{Phys. Rev. Lett.} \textbf{\bibinfo{volume}{98}},
  \bibinfo{pages}{077602} (\bibinfo{year}{2007}).

\bibitem[{\citenamefont{Witzel and Das~Sarma}(2007)}]{Witzel:2007:241303}
\bibinfo{author}{\bibfnamefont{W.~M.} \bibnamefont{Witzel}} \bibnamefont{and}
  \bibinfo{author}{\bibfnamefont{S.}~\bibnamefont{Das~Sarma}},
  \bibinfo{journal}{Phys. Rev. B} \textbf{\bibinfo{volume}{76}},
  \bibinfo{pages}{241303(R)} (\bibinfo{year}{2007}).

\bibitem[{\citenamefont{Zhang et~al.}(2007)\citenamefont{Zhang, Dobrovitski,
  Santos, Viola, and Harmon}}]{Zhang:2007:201302}
\bibinfo{author}{\bibfnamefont{W.~X.} \bibnamefont{Zhang}},
  \bibinfo{author}{\bibfnamefont{V.~V.} \bibnamefont{Dobrovitski}},
  \bibinfo{author}{\bibfnamefont{L.~F.} \bibnamefont{Santos}},
  \bibinfo{author}{\bibfnamefont{L.}~\bibnamefont{Viola}}, \bibnamefont{and}
  \bibinfo{author}{\bibfnamefont{B.~N.} \bibnamefont{Harmon}},
  \bibinfo{journal}{Phys. Rev. B} \textbf{\bibinfo{volume}{75}},
  \bibinfo{pages}{201302(R)} (\bibinfo{year}{2007}).

\bibitem[{\citenamefont{Peng et~al.}(2011)\citenamefont{Peng, Suter, and
  Lidar}}]{Peng:2011:154003}
\bibinfo{author}{\bibfnamefont{X.}~\bibnamefont{Peng}},
  \bibinfo{author}{\bibfnamefont{D.}~\bibnamefont{Suter}}, \bibnamefont{and}
  \bibinfo{author}{\bibfnamefont{D.~A.} \bibnamefont{Lidar}},
  \bibinfo{journal}{J. Phys. B: At. Mol. Opt. Phys.}
  \textbf{\bibinfo{volume}{44}}, \bibinfo{pages}{154003}
  (\bibinfo{year}{2011}).

\bibitem[{\citenamefont{\'Alvarez et~al.}(2010)\citenamefont{\'Alvarez, Ajoy,
  Peng, and Suter}}]{Alvarez:2010:042306}
\bibinfo{author}{\bibfnamefont{G.~A.} \bibnamefont{\'Alvarez}},
  \bibinfo{author}{\bibfnamefont{A.}~\bibnamefont{Ajoy}},
  \bibinfo{author}{\bibfnamefont{X.}~\bibnamefont{Peng}}, \bibnamefont{and}
  \bibinfo{author}{\bibfnamefont{D.}~\bibnamefont{Suter}},
  \bibinfo{journal}{Phys. Rev. A} \textbf{\bibinfo{volume}{82}},
  \bibinfo{pages}{042306} (\bibinfo{year}{2010}).

\bibitem[{\citenamefont{Tyryshkin et~al.}()\citenamefont{Tyryshkin, Wang,
  Zhang, Haller, Ager, Dobrovitski, and Lyon}}]{Tyryshkin:1011.1903}
\bibinfo{author}{\bibfnamefont{A.~M.} \bibnamefont{Tyryshkin}},
  \bibinfo{author}{\bibfnamefont{Z.-H.} \bibnamefont{Wang}},
  \bibinfo{author}{\bibfnamefont{W.}~\bibnamefont{Zhang}},
  \bibinfo{author}{\bibfnamefont{E.~E.} \bibnamefont{Haller}},
  \bibinfo{author}{\bibfnamefont{J.~W.} \bibnamefont{Ager}},
  \bibinfo{author}{\bibfnamefont{V.~V.} \bibnamefont{Dobrovitski}},
  \bibnamefont{and} \bibinfo{author}{\bibfnamefont{S.~A.} \bibnamefont{Lyon}},
  \bibinfo{note}{arXiv:1011.1903v2}.

\bibitem[{\citenamefont{Wang et~al.}(2012)\citenamefont{Wang, Zhang, Tyryshkin,
  Lyon, Ager, Haller, and Dobrovitski}}]{Wang:1011.6417}
\bibinfo{author}{\bibfnamefont{Z.-H.} \bibnamefont{Wang}},
  \bibinfo{author}{\bibfnamefont{W.}~\bibnamefont{Zhang}},
  \bibinfo{author}{\bibfnamefont{A.~M.} \bibnamefont{Tyryshkin}},
  \bibinfo{author}{\bibfnamefont{S.~A.} \bibnamefont{Lyon}},
  \bibinfo{author}{\bibfnamefont{J.~W.} \bibnamefont{Ager}},
  \bibinfo{author}{\bibfnamefont{E.~E.} \bibnamefont{Haller}},
  \bibnamefont{and} \bibinfo{author}{\bibfnamefont{V.~V.}
  \bibnamefont{Dobrovitski}}, \bibinfo{journal}{Phys. Rev. B}
  \textbf{\bibinfo{volume}{85}}, \bibinfo{pages}{085206}
  (\bibinfo{year}{2012}).

\bibitem[{\citenamefont{Barthel et~al.}(2010)\citenamefont{Barthel, Medford,
  Marcus, Hanson, and Gossard}}]{Barthel:2010:266808}
\bibinfo{author}{\bibfnamefont{C.}~\bibnamefont{Barthel}},
  \bibinfo{author}{\bibfnamefont{J.}~\bibnamefont{Medford}},
  \bibinfo{author}{\bibfnamefont{C.~M.} \bibnamefont{Marcus}},
  \bibinfo{author}{\bibfnamefont{M.~P.} \bibnamefont{Hanson}},
  \bibnamefont{and} \bibinfo{author}{\bibfnamefont{A.~C.}
  \bibnamefont{Gossard}}, \bibinfo{journal}{Phys. Rev. Lett.}
  \textbf{\bibinfo{volume}{105}}, \bibinfo{pages}{266808}
  (\bibinfo{year}{2010}).

\bibitem[{\citenamefont{Santos and Viola}(2008)}]{Santos:2008:083009}
\bibinfo{author}{\bibfnamefont{L.~F.} \bibnamefont{Santos}} \bibnamefont{and}
  \bibinfo{author}{\bibfnamefont{L.}~\bibnamefont{Viola}},
  \bibinfo{journal}{New J. Phys.} \textbf{\bibinfo{volume}{10}},
  \bibinfo{pages}{083009} (\bibinfo{year}{2008}).

\bibitem[{\citenamefont{Wang and Liu}(2011{\natexlab{a}})}]{Wang:2011:022306}
\bibinfo{author}{\bibfnamefont{Z.-Y.} \bibnamefont{Wang}} \bibnamefont{and}
  \bibinfo{author}{\bibfnamefont{R.-B.} \bibnamefont{Liu}},
  \bibinfo{journal}{Phys. Rev. A} \textbf{\bibinfo{volume}{83}},
  \bibinfo{pages}{022306} (\bibinfo{year}{2011}{\natexlab{a}}).

\bibitem[{\citenamefont{Uhrig}(2007)}]{Uhrig:2007:100504}
\bibinfo{author}{\bibfnamefont{G.~S.} \bibnamefont{Uhrig}},
  \bibinfo{journal}{Phys. Rev. Lett.} \textbf{\bibinfo{volume}{98}},
  \bibinfo{pages}{100504} (\bibinfo{year}{2007}), \bibinfo{note}{{\it ibid},
  {\textbf{106}}, 129901 (2011)}.

\bibitem[{\citenamefont{Lee et~al.}(2008)\citenamefont{Lee, Witzel, and
  Das~Sarma}}]{Lee:2008:160505}
\bibinfo{author}{\bibfnamefont{B.}~\bibnamefont{Lee}},
  \bibinfo{author}{\bibfnamefont{W.~M.} \bibnamefont{Witzel}},
  \bibnamefont{and}
  \bibinfo{author}{\bibfnamefont{S.}~\bibnamefont{Das~Sarma}},
  \bibinfo{journal}{Phys. Rev. Lett.} \textbf{\bibinfo{volume}{100}},
  \bibinfo{pages}{160505} (\bibinfo{year}{2008}).

\bibitem[{\citenamefont{Uhrig}(2008)}]{Uhrig:2008:083024}
\bibinfo{author}{\bibfnamefont{G.~S.} \bibnamefont{Uhrig}},
  \bibinfo{journal}{New J. Phys.} \textbf{\bibinfo{volume}{10}},
  \bibinfo{pages}{083024} (\bibinfo{year}{2008}).

\bibitem[{\citenamefont{Yang and Liu}(2008)}]{Yang:2008:180403}
\bibinfo{author}{\bibfnamefont{W.}~\bibnamefont{Yang}} \bibnamefont{and}
  \bibinfo{author}{\bibfnamefont{R.-B.} \bibnamefont{Liu}},
  \bibinfo{journal}{Phys. Rev. Lett.} \textbf{\bibinfo{volume}{101}},
  \bibinfo{pages}{180403} (\bibinfo{year}{2008}).

\bibitem[{\citenamefont{Uhrig and Lidar}(2010)}]{Uhrig:2010:012301}
\bibinfo{author}{\bibfnamefont{G.~S.} \bibnamefont{Uhrig}} \bibnamefont{and}
  \bibinfo{author}{\bibfnamefont{D.~A.} \bibnamefont{Lidar}},
  \bibinfo{journal}{Phys. Rev. A} \textbf{\bibinfo{volume}{82}},
  \bibinfo{pages}{012301} (\bibinfo{year}{2010}).

\bibitem[{\citenamefont{Pasini et~al.}(2008)\citenamefont{Pasini, Fischer,
  Karbach, and Uhrig}}]{Pasini:2008:032315}
\bibinfo{author}{\bibfnamefont{S.}~\bibnamefont{Pasini}},
  \bibinfo{author}{\bibfnamefont{T.}~\bibnamefont{Fischer}},
  \bibinfo{author}{\bibfnamefont{P.}~\bibnamefont{Karbach}}, \bibnamefont{and}
  \bibinfo{author}{\bibfnamefont{G.~S.} \bibnamefont{Uhrig}},
  \bibinfo{journal}{Phys. Rev. A} \textbf{\bibinfo{volume}{77}},
  \bibinfo{pages}{032315} (\bibinfo{year}{2008}).

\bibitem[{\citenamefont{Fauseweh et~al.}(2012)\citenamefont{Fauseweh, Pasini,
  and Uhrig}}]{Fauseweh:1112.0446}
\bibinfo{author}{\bibfnamefont{B.}~\bibnamefont{Fauseweh}},
  \bibinfo{author}{\bibfnamefont{S.}~\bibnamefont{Pasini}}, \bibnamefont{and}
  \bibinfo{author}{\bibfnamefont{G.~S.} \bibnamefont{Uhrig}},
  \bibinfo{journal}{Phys. Rev. A} \textbf{\bibinfo{volume}{85}},
  \bibinfo{pages}{022310} (\bibinfo{year}{2012}).

\bibitem[{\citenamefont{Biercuk
  et~al.}(2009{\natexlab{a}})\citenamefont{Biercuk, Uys, VanDevender, Shiga,
  Itano, and Bollinger}}]{Biercuk:2009:996}
\bibinfo{author}{\bibfnamefont{M.~J.} \bibnamefont{Biercuk}},
  \bibinfo{author}{\bibfnamefont{H.}~\bibnamefont{Uys}},
  \bibinfo{author}{\bibfnamefont{A.~P.} \bibnamefont{VanDevender}},
  \bibinfo{author}{\bibfnamefont{N.}~\bibnamefont{Shiga}},
  \bibinfo{author}{\bibfnamefont{W.~M.} \bibnamefont{Itano}}, \bibnamefont{and}
  \bibinfo{author}{\bibfnamefont{J.~J.} \bibnamefont{Bollinger}},
  \bibinfo{journal}{Nature} \textbf{\bibinfo{volume}{458}},
  \bibinfo{pages}{996} (\bibinfo{year}{2009}{\natexlab{a}}).

\bibitem[{\citenamefont{Du et~al.}(2009)\citenamefont{Du, Rong, Zhao, Wang,
  Yang, and Liu}}]{Du:2009:1265}
\bibinfo{author}{\bibfnamefont{J.}~\bibnamefont{Du}},
  \bibinfo{author}{\bibfnamefont{X.}~\bibnamefont{Rong}},
  \bibinfo{author}{\bibfnamefont{N.}~\bibnamefont{Zhao}},
  \bibinfo{author}{\bibfnamefont{Y.}~\bibnamefont{Wang}},
  \bibinfo{author}{\bibfnamefont{J.}~\bibnamefont{Yang}}, \bibnamefont{and}
  \bibinfo{author}{\bibfnamefont{R.-B.} \bibnamefont{Liu}},
  \bibinfo{journal}{Nature} \textbf{\bibinfo{volume}{461}},
  \bibinfo{pages}{1265} (\bibinfo{year}{2009}).

\bibitem[{\citenamefont{Biercuk
  et~al.}(2009{\natexlab{b}})\citenamefont{Biercuk, Uys, VanDevender, Shiga,
  Itano, and Bollinger}}]{Michael:2009:062324}
\bibinfo{author}{\bibfnamefont{M.~J.} \bibnamefont{Biercuk}},
  \bibinfo{author}{\bibfnamefont{H.}~\bibnamefont{Uys}},
  \bibinfo{author}{\bibfnamefont{A.~P.} \bibnamefont{VanDevender}},
  \bibinfo{author}{\bibfnamefont{N.}~\bibnamefont{Shiga}},
  \bibinfo{author}{\bibfnamefont{W.~M.} \bibnamefont{Itano}}, \bibnamefont{and}
  \bibinfo{author}{\bibfnamefont{J.~J.} \bibnamefont{Bollinger}},
  \bibinfo{journal}{Phys. Rev. A} \textbf{\bibinfo{volume}{79}},
  \bibinfo{pages}{062324} (\bibinfo{year}{2009}{\natexlab{b}}).

\bibitem[{\citenamefont{Uys et~al.}(2009)\citenamefont{Uys, Biercuk, and
  Bollinger}}]{Uys:2009:040501}
\bibinfo{author}{\bibfnamefont{H.}~\bibnamefont{Uys}},
  \bibinfo{author}{\bibfnamefont{M.~J.} \bibnamefont{Biercuk}},
  \bibnamefont{and} \bibinfo{author}{\bibfnamefont{J.~J.}
  \bibnamefont{Bollinger}}, \bibinfo{journal}{Phys. Rev. Lett.}
  \textbf{\bibinfo{volume}{103}}, \bibinfo{pages}{040501}
  (\bibinfo{year}{2009}).

\bibitem[{\citenamefont{Jenista et~al.}(2009)\citenamefont{Jenista, Stokes,
  Branca, and Warren}}]{Jenista:2009:204510}
\bibinfo{author}{\bibfnamefont{E.~R.} \bibnamefont{Jenista}},
  \bibinfo{author}{\bibfnamefont{A.~M.} \bibnamefont{Stokes}},
  \bibinfo{author}{\bibfnamefont{R.~T.} \bibnamefont{Branca}},
  \bibnamefont{and} \bibinfo{author}{\bibfnamefont{W.~S.}
  \bibnamefont{Warren}}, \bibinfo{journal}{J. Chem. Phys.}
  \textbf{\bibinfo{volume}{131}}, \bibinfo{pages}{204510}
  (\bibinfo{year}{2009}).

\bibitem[{\citenamefont{Uhrig}(2009)}]{Uhrig:2009:120502}
\bibinfo{author}{\bibfnamefont{G.~S.} \bibnamefont{Uhrig}},
  \bibinfo{journal}{Phys. Rev. Lett.} \textbf{\bibinfo{volume}{102}},
  \bibinfo{pages}{120502} (\bibinfo{year}{2009}).

\bibitem[{\citenamefont{West et~al.}(2010)\citenamefont{West, Fong, and
  Lidar}}]{West:2010:130501}
\bibinfo{author}{\bibfnamefont{J.~R.} \bibnamefont{West}},
  \bibinfo{author}{\bibfnamefont{B.~H.} \bibnamefont{Fong}}, \bibnamefont{and}
  \bibinfo{author}{\bibfnamefont{D.~A.} \bibnamefont{Lidar}},
  \bibinfo{journal}{Phys. Rev. Lett.} \textbf{\bibinfo{volume}{104}},
  \bibinfo{pages}{130501} (\bibinfo{year}{2010}).

\bibitem[{\citenamefont{Quiroz and Lidar}(2011)}]{Quiroz:1105.4303}
\bibinfo{author}{\bibfnamefont{G.}~\bibnamefont{Quiroz}} \bibnamefont{and}
  \bibinfo{author}{\bibfnamefont{D.~A.} \bibnamefont{Lidar}},
  \bibinfo{journal}{Phys. Rev. A} \textbf{\bibinfo{volume}{84}},
  \bibinfo{pages}{042328} (\bibinfo{year}{2011}).

\bibitem[{\citenamefont{Kuo and Lidar}(2011)}]{Kuo:1106.2151}
\bibinfo{author}{\bibfnamefont{W.-J.} \bibnamefont{Kuo}} \bibnamefont{and}
  \bibinfo{author}{\bibfnamefont{D.~A.} \bibnamefont{Lidar}},
  \bibinfo{journal}{Phys. Rev. A} \textbf{\bibinfo{volume}{84}},
  \bibinfo{pages}{042329} (\bibinfo{year}{2011}).

\bibitem[{\citenamefont{Jiang and Imambekov}(2011)}]{Jiang:2011:060302}
\bibinfo{author}{\bibfnamefont{L.}~\bibnamefont{Jiang}} \bibnamefont{and}
  \bibinfo{author}{\bibfnamefont{A.}~\bibnamefont{Imambekov}},
  \bibinfo{journal}{Phys. Rev. A} \textbf{\bibinfo{volume}{84}},
  \bibinfo{pages}{060302} (\bibinfo{year}{2011}).

\bibitem[{\citenamefont{Mukhtar et~al.}(2010)\citenamefont{Mukhtar, Saw, Soh,
  and Gong}}]{Mukhtar:2010:012331}
\bibinfo{author}{\bibfnamefont{M.}~\bibnamefont{Mukhtar}},
  \bibinfo{author}{\bibfnamefont{T.~B.} \bibnamefont{Saw}},
  \bibinfo{author}{\bibfnamefont{W.~T.} \bibnamefont{Soh}}, \bibnamefont{and}
  \bibinfo{author}{\bibfnamefont{J.}~\bibnamefont{Gong}},
  \bibinfo{journal}{Phys. Rev. A} \textbf{\bibinfo{volume}{81}},
  \bibinfo{pages}{012331} (\bibinfo{year}{2010}).

\bibitem[{\citenamefont{Wang and Liu}(2011{\natexlab{b}})}]{Wang:2011:062313}
\bibinfo{author}{\bibfnamefont{Z.-Y.} \bibnamefont{Wang}} \bibnamefont{and}
  \bibinfo{author}{\bibfnamefont{R.-B.} \bibnamefont{Liu}},
  \bibinfo{journal}{Phys. Rev. A} \textbf{\bibinfo{volume}{83}},
  \bibinfo{pages}{062313} (\bibinfo{year}{2011}{\natexlab{b}}).

\bibitem[{\citenamefont{Cywi\'{n}ski et~al.}(2008)\citenamefont{Cywi\'{n}ski,
  Lutchyn, Nave, and Das~Sarma}}]{Cywinski:2008:174509}
\bibinfo{author}{\bibfnamefont{{\L}.}~\bibnamefont{Cywi\'{n}ski}},
  \bibinfo{author}{\bibfnamefont{R.~M.} \bibnamefont{Lutchyn}},
  \bibinfo{author}{\bibfnamefont{C.~P.} \bibnamefont{Nave}}, \bibnamefont{and}
  \bibinfo{author}{\bibfnamefont{S.}~\bibnamefont{Das~Sarma}},
  \bibinfo{journal}{Phys. Rev. B} \textbf{\bibinfo{volume}{77}},
  \bibinfo{pages}{174509} (\bibinfo{year}{2008}).

\bibitem[{\citenamefont{Hodgson et~al.}(2010)\citenamefont{Hodgson, Viola, and
  D'Amico}}]{Hodgson:2010:062321}
\bibinfo{author}{\bibfnamefont{T.~E.} \bibnamefont{Hodgson}},
  \bibinfo{author}{\bibfnamefont{L.}~\bibnamefont{Viola}}, \bibnamefont{and}
  \bibinfo{author}{\bibfnamefont{I.}~\bibnamefont{D'Amico}},
  \bibinfo{journal}{Phys. Rev. A} \textbf{\bibinfo{volume}{81}},
  \bibinfo{pages}{062321} (\bibinfo{year}{2010}).

\bibitem[{\citenamefont{Khodjasteh et~al.}(2011)\citenamefont{Khodjasteh,
  Erd\'elyi, and Viola}}]{Khodjasteh:2011:020305}
\bibinfo{author}{\bibfnamefont{K.}~\bibnamefont{Khodjasteh}},
  \bibinfo{author}{\bibfnamefont{T.}~\bibnamefont{Erd\'elyi}},
  \bibnamefont{and} \bibinfo{author}{\bibfnamefont{L.}~\bibnamefont{Viola}},
  \bibinfo{journal}{Phys. Rev. A} \textbf{\bibinfo{volume}{83}},
  \bibinfo{pages}{020305} (\bibinfo{year}{2011}).

\bibitem[{\citenamefont{Ajoy et~al.}(2011)\citenamefont{Ajoy, \'Alvarez, and
  Suter}}]{Ajoy:2011:032303}
\bibinfo{author}{\bibfnamefont{A.}~\bibnamefont{Ajoy}},
  \bibinfo{author}{\bibfnamefont{G.~A.} \bibnamefont{\'Alvarez}},
  \bibnamefont{and} \bibinfo{author}{\bibfnamefont{D.}~\bibnamefont{Suter}},
  \bibinfo{journal}{Phys. Rev. A} \textbf{\bibinfo{volume}{83}},
  \bibinfo{pages}{032303} (\bibinfo{year}{2011}).

\bibitem[{\citenamefont{Pasini and Uhrig}(2010)}]{Pasini:2010:012309}
\bibinfo{author}{\bibfnamefont{S.}~\bibnamefont{Pasini}} \bibnamefont{and}
  \bibinfo{author}{\bibfnamefont{G.~S.} \bibnamefont{Uhrig}},
  \bibinfo{journal}{Phys. Rev. A} \textbf{\bibinfo{volume}{81}},
  \bibinfo{pages}{012309} (\bibinfo{year}{2010}).

\bibitem[{\citenamefont{Chen and Liu}(2010)}]{Chen:2010:052324}
\bibinfo{author}{\bibfnamefont{K.}~\bibnamefont{Chen}} \bibnamefont{and}
  \bibinfo{author}{\bibfnamefont{R.-B.} \bibnamefont{Liu}},
  \bibinfo{journal}{Phys. Rev. A} \textbf{\bibinfo{volume}{82}},
  \bibinfo{pages}{052324} (\bibinfo{year}{2010}).

\bibitem[{\citenamefont{Gordon et~al.}(2008)\citenamefont{Gordon, Kurizki, and
  Lidar}}]{Gordon:2008:010403}
\bibinfo{author}{\bibfnamefont{G.}~\bibnamefont{Gordon}},
  \bibinfo{author}{\bibfnamefont{G.}~\bibnamefont{Kurizki}}, \bibnamefont{and}
  \bibinfo{author}{\bibfnamefont{D.~A.} \bibnamefont{Lidar}},
  \bibinfo{journal}{Phys. Rev. Lett.} \textbf{\bibinfo{volume}{101}},
  \bibinfo{pages}{010403} (\bibinfo{year}{2008}).

\bibitem[{\citenamefont{Anderson}(1954)}]{Anderson:1954:316}
\bibinfo{author}{\bibfnamefont{P.~W.} \bibnamefont{Anderson}},
  \bibinfo{journal}{J. Phys. Soc. Jpn.} \textbf{\bibinfo{volume}{9}},
  \bibinfo{pages}{316} (\bibinfo{year}{1954}).

\bibitem[{\citenamefont{Kubo}(1954)}]{Kubo:1954:935}
\bibinfo{author}{\bibfnamefont{R.}~\bibnamefont{Kubo}}, \bibinfo{journal}{J.
  Phys. Soc. Jpn.} \textbf{\bibinfo{volume}{9}}, \bibinfo{pages}{935}
  (\bibinfo{year}{1954}).

\bibitem[{\citenamefont{Uhrig and Pasini}(2010)}]{Uhrig:2010:045001}
\bibinfo{author}{\bibfnamefont{G.~S.} \bibnamefont{Uhrig}} \bibnamefont{and}
  \bibinfo{author}{\bibfnamefont{S.}~\bibnamefont{Pasini}},
  \bibinfo{journal}{New J. Phys.} \textbf{\bibinfo{volume}{12}},
  \bibinfo{pages}{045001} (\bibinfo{year}{2010}).

\bibitem[{\citenamefont{Wang and Liu}()}]{Wang:unpublished}
\bibinfo{author}{\bibfnamefont{Z.-Y.} \bibnamefont{Wang}} \bibnamefont{and}
  \bibinfo{author}{\bibfnamefont{R.-B.} \bibnamefont{Liu}},
  \bibinfo{note}{unpublished}.

\bibitem[{\citenamefont{Gordon et~al.}(2006)\citenamefont{Gordon, Kurizki, and
  Kofman}}]{Gordon:2006:398}
\bibinfo{author}{\bibfnamefont{G.}~\bibnamefont{Gordon}},
  \bibinfo{author}{\bibfnamefont{G.}~\bibnamefont{Kurizki}}, \bibnamefont{and}
  \bibinfo{author}{\bibfnamefont{A.~G.} \bibnamefont{Kofman}},
  \bibinfo{journal}{Opt. Commun.} \textbf{\bibinfo{volume}{264}},
  \bibinfo{pages}{398} (\bibinfo{year}{2006}).

\bibitem[{\citenamefont{Gordon and Kurizki}(2007)}]{Gordon:2007:042310}
\bibinfo{author}{\bibfnamefont{G.}~\bibnamefont{Gordon}} \bibnamefont{and}
  \bibinfo{author}{\bibfnamefont{G.}~\bibnamefont{Kurizki}},
  \bibinfo{journal}{Phys. Rev. A} \textbf{\bibinfo{volume}{76}},
  \bibinfo{pages}{042310} (\bibinfo{year}{2007}).

\bibitem[{\citenamefont{Gordon et~al.}(2007)\citenamefont{Gordon, Erez, and
  Kurizki}}]{Gordon:2007:S75}
\bibinfo{author}{\bibfnamefont{G.}~\bibnamefont{Gordon}},
  \bibinfo{author}{\bibfnamefont{N.}~\bibnamefont{Erez}}, \bibnamefont{and}
  \bibinfo{author}{\bibfnamefont{G.}~\bibnamefont{Kurizki}},
  \bibinfo{journal}{J. Phys. B} \textbf{\bibinfo{volume}{40}},
  \bibinfo{pages}{S75} (\bibinfo{year}{2007}).

\bibitem[{\citenamefont{Berman and Brewer}(1985)}]{Berman:1985:2784}
\bibinfo{author}{\bibfnamefont{P.~R.} \bibnamefont{Berman}} \bibnamefont{and}
  \bibinfo{author}{\bibfnamefont{R.~G.} \bibnamefont{Brewer}},
  \bibinfo{journal}{Phys. Rev. A} \textbf{\bibinfo{volume}{32}},
  \bibinfo{pages}{2784} (\bibinfo{year}{1985}).

\bibitem[{\citenamefont{Sun and Liu}(1996)}]{Sun:1996:343}
\bibinfo{author}{\bibfnamefont{C.~P.} \bibnamefont{Sun}} \bibnamefont{and}
  \bibinfo{author}{\bibfnamefont{X.~J.} \bibnamefont{Liu}},
  \bibinfo{journal}{Acta Phys. Sin.} \textbf{\bibinfo{volume}{5}},
  \bibinfo{pages}{343} (\bibinfo{year}{1996}).

\end{thebibliography}

\end{document}